\documentclass[article]{aa}
\usepackage{epsfig}
\usepackage{graphicx}
\begin{document}
 
                          
\title{Kinematics of the Local Universe XIII.}
\subtitle{21-cm line measurements of 452 galaxies 
with the Nan\c cay radiotelescope, JHK Tully-Fisher relation 
and preliminary maps of the peculiar velocity field}

\author{G.\ Theureau\inst{1,2} \and M.\ 0.\ Hanski\inst{1,3} \and N.\ Coudreau\inst{2} \and N.\ Hallet\inst{2} \and 
J.-M.\ Martin\inst{2}}
\offprints{G.Theureau, \email{theureau@cnrs-orleans.fr}}
\institute{ LPCE/CNRS UMR6115, F45071 Orléans Cedex 02, France
\and
Observatoire de Paris/Meudon, GEPI/CNRS URA1757, F92195 Meudon
Principal Cedex, France
\and
Tuorla observatory, University of Turku, SF 21500 Piikki\"o, Finland
}
\date{Received : August 8th, 2006 \hspace*{3cm}; accepted : November 3rd, 2006}

\authorrunning{Theureau et al.}
\titlerunning{Kinematics of the Local Universe XIII}

\abstract{}{This paper presents 452 new 21-cm neutral hydrogen line 
measurements carried out with the FORT receiver of the meridian 
transit Nan\c cay radiotelescope (NRT) in the period April 2003 -- March 2005.}{ 
This observational programme is part of a larger project aiming at 
collecting an exhaustive and magnitude-complete HI extragalactic 
catalogue for Tully-Fisher applications (the so-called KLUN project, 
for Kinematics of the Local Universe studies, end in 2008).
The whole on-line HI archive of the NRT contains today reduced HI-profiles for ~4500
spiral galaxies of declination $\delta > -40^{\circ}$ 
(http://klun.obs-nancay.fr).}{
As an example of application, we use direct Tully-Fisher relation in three (JHK) bands  
in deriving distances to a large catalog of 3126 spiral galaxies distributed through 
the whole sky and sampling well the radial velocity range between 0 and 8\,000~km~s$^{-1}$.
Thanks to an iterative method accounting for selection bias and smoothing effects,
we show as a preliminary output a detailed and original map of the velocity field in 
the Local Universe.}{}


\maketitle

\section{Introduction}

The present paper complements the KLUN\footnote{for Kinematics in the Local UNiverse} 
data-series (I: Bottinelli et al.\, \cite{bot92};
II: Bottinelli et al.\, \cite{bot93}; III: di Nella et al.\, \cite{nel96}, 
VII: Theureau et al.\, \cite{the98a}, XII: Paturel et al.\, \cite{pat03b},
 Theureau et al.\, \cite{the05})
with a collection of HI line measurements acquired with the Nan\c cay radiotelescope
(FORT)\footnote{data tables and HI-profiles
and corresponding comments are available in electronic form
at the CDS via anonymous ftp to cdsarc.u-strasbg.fr (130.79.128.5)
or via http://cdsweb.u-strasbg.fr/Abstract.html, or directly
at our web site http://klun.obs-nancay.fr} . 
This programme has received the label of key project of the instrument
and is allocated on average 20 \% of observing time since the first light in mid 2000.

The input catalogue has been carried out from a compilation of the HYPERLEDA extragalactic database
completed by the 2.7 million galaxy catalogue extracted from the DSS (Paturel et al.\, \cite{pat00}), and 
the releases of DENIS (DEep Near Infrared Survey, Mamon et al, \cite{mam04}) and 2MASS (2 
Micron All Sky Survey, Jarret et al, \cite{jar00}) near infrared CCD surveys.
The aim of the programme is to build a large all sky sample of spiral galaxies, 
complete down to well defined magnitude limits in the five photometric bands 
$B$, $I$, $J$, $H$ and $K$, and to allow peculiar velocity mapping of galaxies
up to 10,000 km.s$^{-1}$ in radial velocity, i.e. up to a scale greater than the largest
structures of the Local Universe.

This programme is complementary to other large HI projects such as 
HIPASS\footnote{http://www.atnf.csiro.au/research/multibeam/release/} in Parkes 
(Barnes et al \cite{bar01}) or 
the ALFA-project at ARECIBO\footnote{http://alfa.naic.edu/alfa/}. 
The majority of the objects we observed from Nan\c cay are in the range 
(-40$^{\circ}$,+0$^{\circ}$) in declination, thus favouring the declination 
range unreachable by ARECIBO. Our aim was to fill the gaps left in the last
hyperleda HI compilation by Paturel et al.\ (\cite{pat03b}) in order to reach
well defined selection criteria in terms of redshift coverage and magnitude 
completeness (see Sect. 3.).

This kind of HI data is crucial for constraining the gas and total mass function of spiral galaxies
as a function of morphology and environnement, it allows also the mapping of the total mass distribution
from peculiar velocities and thus provides strong constraints on cosmological models and large 
scale structure formation. They can in particular provide a unique starting point
for total mass power spectrum studies.

Study of peculiar velocities allows the verification of the current
theory of cosmological structure formation by gravitational instability.
It gives information on bulk motion, and the value of $\Omega_{\rm m}$ (cf.\ reviews by
Willick \cite{wil00} and Zaroubi \cite{zar02}, and the comprehensive work by Strauss
\& Willick \cite{str95}).  The velocity measurements are done using redshift
independent secondary distance indicators, such as the Tully-Fisher (TF)
relation for spiral galaxies, the Faber-Jackson, $D_{\rm n}$-$\sigma$, Fundamental Plane (FP) relation,
or the surface brightness fluctuations for early type galaxies.
The largest surveys so far are the Mathewson and Ford \cite{mat96} sample, the MARK III (Willick et al.  
\cite{wil97}), SFI (Giovanelli et al. \cite{gio97}, Haynes et al.  \cite{hay99}), 
ENEAR (da Costa et al. \cite{dac00a},\cite{dac00b}), and the updated FGC
catalogue (2MFGC, Mitronova et al. \cite{mit04}).
Each contains in the order of one or two thousand independent distance estimates
in the local 80~$h^{-1}$~Mpc volume.  The use is then to compare them to the velocity field derived from
the galaxy density distribution as infered from complete redshift sample (e.g. PSCz, Saunders et al. \cite{sau00},
or NOG, Marinoni et al. \cite{mar98})
                                                                                                                      
Our own Kinematics of the Local Universe (KLUN) TF sample has been used in the
study of $H_0$ (Theureau et al. \cite{the97}, Ekholm et al. \cite{ekh99})
and local structures (Hanski et al. \cite{han01}). 
The sample consists of all the galaxies with published rotational velocities 
collected in the HYPERLEDA\footnote{http://leda.univ-lyon1.fr},
database (Paturel et al. \cite{pat03a}), plus the recent large KLUN+ contribution 
(Theureau et al 2005 and this paper). The total Tully-Fisher sample counts 15\,700 spirals
and uses five different
wavelength galaxy magnitudes. B- and I-magnitudes come from various sources, carefully
homogenized to a common system. The largest sources are DSS1 for B, and
Mathewson et al.\ (\cite{mat92}, \cite{mat96}) and DENIS (Paturel et al. \cite{pat05}) for I band.
J, H, and K-magnitudes are
from the 2MASS\footnote{http://www.ipac.caltech.edu/2mass/} survey (Jarret et al. \cite{jar00}). 
The 2MASS magnitudes, taken from a single survey, avoid
any problems that the homogenization may cause, and are thus exclusively
used in data analysis.
Further, we exclude the measurements with large errors and the galaxies
that for other reasons, explained later in the text, are unsuitable
for this study. 3126 galaxies remain, which we use
for the mapping of the peculiar
velocity field within the radius of 80~$h^{-1}$~Mpc.

The paper is structured as follows:
the Nan\c cay radiotelescope, the processing chain and the reduced HI data are presented in Sect. \ref{SecObs};
The characteristics of the input Tully-Fisher catalogue are listed in Sect. \ref{SecSam}, while the iterative 
method to obtain unbiased peculiar velocities from it is given in Sect. \ref{SecND};
finally, we give in Sect. \ref{SecRes} some preliminary results and show some 
examples of peculiar velocity map realization.

\begin{figure}
\epsfig{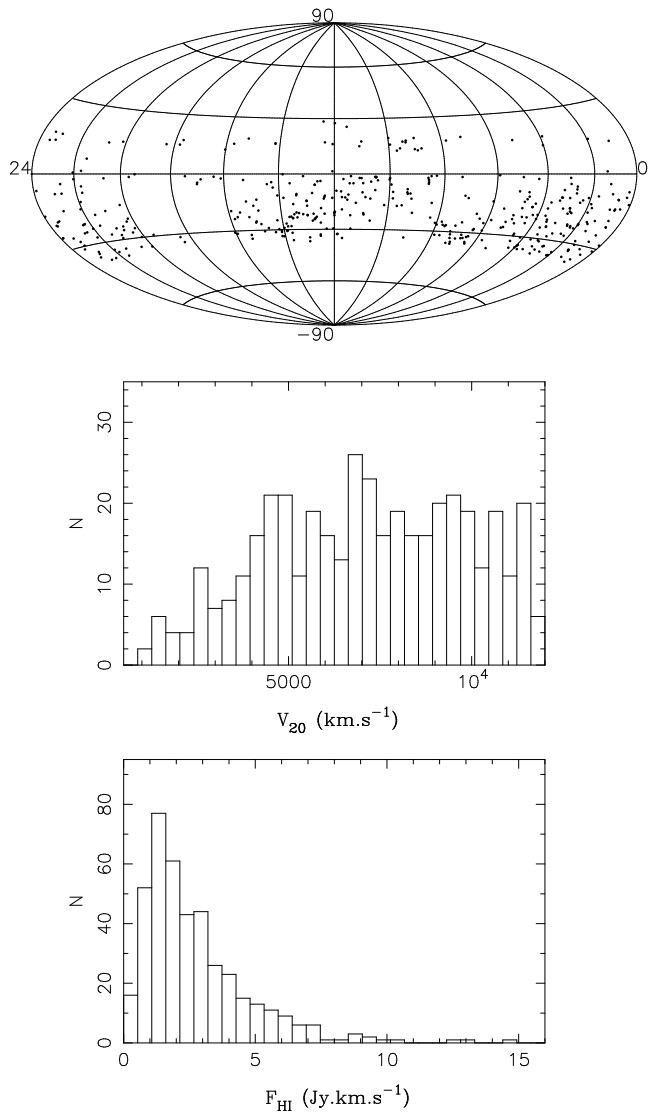}
\caption[]
{Aitoff projection of the observed sample in J2000 equatorial coordinates,
histogram of radial velocities $V_{20}$ and HI fluxes (see Table \ref{TabPar})}
\end{figure}

\section{The HI data \label{SecObs}}

\subsection{The Nan\c cay observations}

The Nan\c cay radiotelescope (France) is a
single dish antenna with a collecting area of 6912 m$^2$ (200
$\times$ 34.56) equivalent to that of a 94 m-diameter parabolic dish.
The half-power beam width at 21-cm is 3.6 arcmin (EW) $\times$ 22 arcmin
(NS) (at zero declination). The minimal system temperature at $\delta =
15^{\circ}$ is about 35 K in both horizontal and vertical polarizations.
The spectrometer is a 8192-channel autocorrelator offering a maximal 
bandwidth of 50 MHz. In this mode, and with two banks in vertical 
and horizontal polarizations counting 4096 channels each, the spacing of 
the channels corresponds to 2.6 km.s$^{-1}$ at 21 cm. After boxcar smoothing the 
final resolution is typically $\sim$10 km.s$^{-1}$. The 50 MHz bandwidth is
centered on $\sim$ 1387 MHz, thus corresponding to an interval of 10,500 km.s$^{-1}$ centered
on a velocity of 7000 km.s$^{-1}$ (except for the few objects with a radial velocity known 
to be less than 2000 km.s$^{-1}$, for which the observing band was centered on 5000 km.s$^{-1}$). 
The relative gain of the antenna has been calibrated according to Fouqu\'e et al.\,
(\cite{fou90}); the final HI-fluxes (Table \ref{TabPar}) are calibrated using as templates 
a set of well-defined radio continuum sources observed each month. 

One "observation" is a series of ON/OFF observational sequences; each sequence
is made of ten elementary integrations of 4 seconds each, plus a set of 
3 integrations of 2 seconds for the calibration, adding up in each cycle to
40+6 seconds for the source and 40+6 seconds for the comparison field.
The comparison field is taken at exactly the same positions of the focal track as the source in the same cycle. 
In this way one minimizes efficiently the difference between ON and OFF total power.
A typical meridian transit observation lasts about 35 minutes and is centered on the meridian,
where the gain is known to be at its maximum; it contains a series of $\sim$20 ON/OFF cycles.
 
The processing chain consists of selecting good elementary integrations or
cycles, masking and interpolating areas in the time-frequency plane, 
straightening the base-line by a polynomial fit (order in the range 1-6), and
applying a boxcar smoothing.
The maximum of the line profile is determined by eye as the mean value
of the maxima of its two horns after taking into account the rms
noise (estimated in the base-line). The widths, measured at the standard
levels of 20\% and 50\% of that maximum, correspond to the "distance"
separating the two external points of the profile at these intensity levels.
The signal to noise ratio is the maximum of the line (see above) over rms noise
in the baseline fitted region.

The total list of corrected HI-astrophysical parameters (Table \ref{TabPar}), 
21-cm line profiles (Fig. 3), 
and comments concerning the profiles (Table \ref{TabCom}), are available in electronic form
at the CDS via anonymous ftp to cdsarc.u-strasbg.fr 
or via http://cdsweb.u-strasbg.fr/Abstract.html. 

\begin{figure}
\epsfig{file=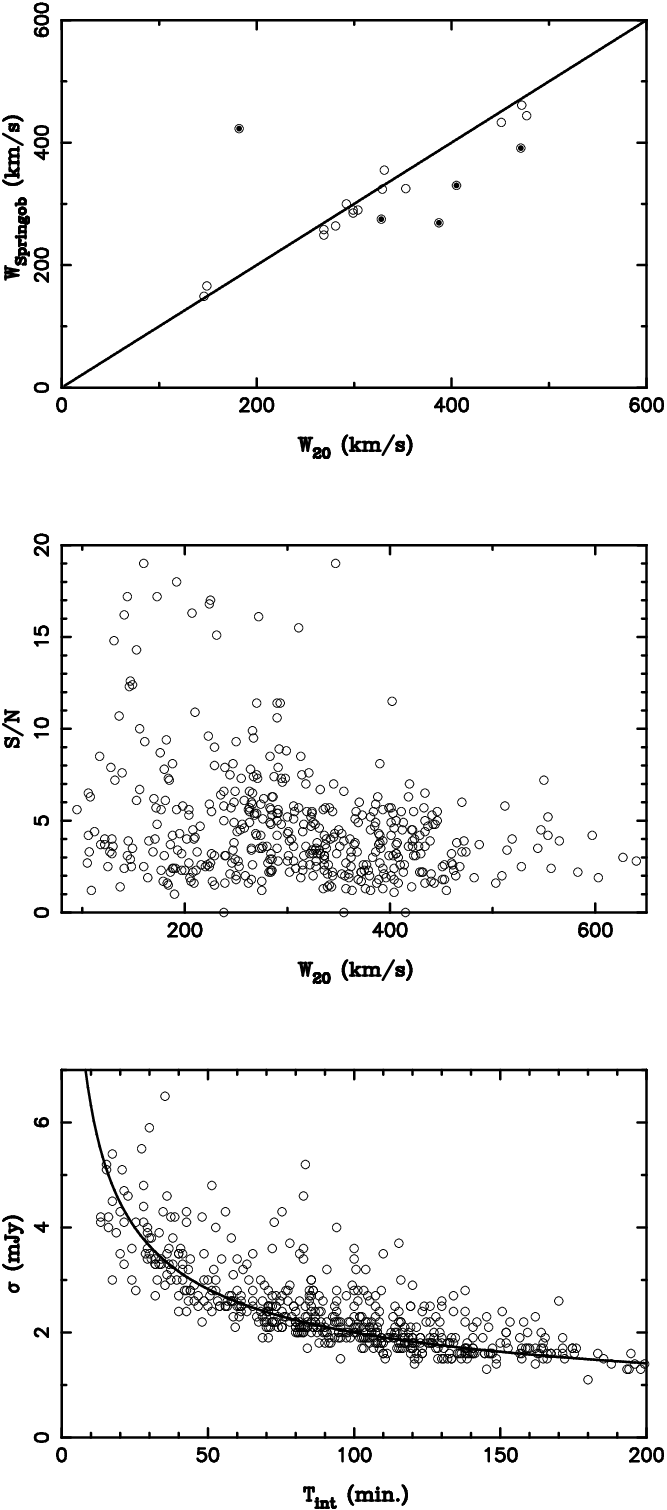,width=8.5cm}
\caption[]
{Top : comparison of some of our HI-line width at the 20\% level with some independent
measurements from Springob et al.  \cite{spr05}.
Middle : distribution of signal to noise ratio $S/N$
as a function of 20\% level line width $W_{20}$.
Bottom :  rms noise $\sigma$ in mJy (outside the 21-cm line)
versus integration time. The curve shows the line $\sigma=20/\sqrt{T_{int}}$
}
\label{FigStat}
\end{figure}

\subsection{Sample characteristics and data description }

In the first five years of observations (2001-2005), since the upgrade of the Nan\c cay receiver (FORT),
we have observed 2500 galaxies, successfully detected about 1600 of them and fully reduced 1340 HI profiles. 

As a second KLUN+ release, we present here the spectra obtained for 452 of these 
galaxies, observed between April 2003 and March 2005. 
Some simple statistics is presented on figure \ref{FigStat}. The upper panel shows
a comparison of some of our HI-line width at the 20\% level with equivalent measurements 
($W_{P20}$) found in the last large compilation of line widths by Springob et al. \cite{spr05}. 
The overlap is quite small, concerning 20 galaxies only. The few outlying galaxies marked
with filled circles are identified either as distorted HI-line, at the limit of detectability 
or HI-confused with another neighbouring galaxy (case of pgc2350, pgc20363, pgc67934, pgc66850 
and pgc54825, see Table \ref{TabCom}). The other ones
are well aligned on the first bisecting line. Anyway, one could eventually guess a small
systematic effect there : a slight over-estimation of the line width for large
widths with respect to H$_{\alpha}$ or Arecibo measurement. This is explained easily 
by the general low signal to noise ratio we have for edge-on galaxies, in the range of
fluxes we are concerned with.  
The Middle panel shows the distribution of signal to noise ratio $S/N$
as a function of 20\% level line width $W_{20}$, and the bottom panel 
shows the rms noise $\sigma$ in mJy (outside the 21-cm line) 
versus integration time. In the latter, the curve shows the line $\sigma=20/\sqrt{T_{int}}$.  

Table \ref{TabPar} contains all the reduced HI parameters. Table \ref{TabCom} provides corresponding
comments, when necessary, for each galaxy. Comments concern mainly object designation, peculiar morphologies
or peculiar HI line shape, spectrum quality and HI confusions. The spectra and extracted data are assigned 
a quality code. A flag '?' or '*' warns for suspected or confirmed 
HI line confusion. The five quality classes are defined as follows:
\begin{itemize}
\item A : high quality spectrum, high signal to noise and well defined HI profile
\item B : good signal to noise ratio, line border well defined, still suitable for Tully-Fisher applications
\item C : low signal to noise, noisy or asymmetrical profile, well detected but one 
should not trust the line width. The radial velocity is perfectly determined
\item D : low signal to noise, noisy profile at the limit of detection. Probably detected, 
but even radial velocity could be doubtful
\item E : not detected. 
The absence of detection, corresponding to the "E" code in the notes, is due to several
possible reasons: either the object was too faint in HI to be detected within a reasonable
integration time (120 full ON/OFF cycles, equivalent to 3 meridian transit), which is probably
the major case, or we did not know its radial velocity and it fell outside the frequency range,
or the HI line was always behind a radar emission or an interference... In a few cases, some standing
waves are clearly visible in the full bandwidth plots (50 MHz $\equiv$ 10,500 km.s$^{-1}$). These are 
due to reflexions either in the cables, between the primary and secondary mirrors or between the 
secondary and tertiary mirrors. It happens when a strong radiosource (often the Sun) is close to the 
main beam of the antenna.
Finally, when "no detection" is stated, the line was expected to fall within the
observed frequency band and the value of the noise gives a fair upper limit for the HI signal.
\end{itemize}

The distribution of the targets among the different classes is summarized in table \ref{TabQ}.

\begin{table}
\caption{Statistics of the detected galaxies vs. HI profile class. Among the "C" galaxies, 38 were flagged
as HI confused}
\begin{tabular}{cl}
\hline
 profile class & nb of galaxies \\
\hline
A &  84  \\
B & 179  \\
C & 161  \\
D &  28  \\
E & 101  \\
\hline
\end{tabular}
\label{TabQ}
\end{table}

The Aitoff projection of the catalogue in J2000 equatorial is seen on Fig.\ 1, together with the 
distribution of the radial velocities and HI fluxes. 
Most of the observed objects are in the range 4,000-10,000 km s$^{-1}$
where the lack of Tully-Fisher measurements in the literature is the most critical.  

\subsection{Data description \label{SecDat}}

\paragraph{Radial velocities \label{RadVel}}

Our observed radial velocities are listed in Table \ref{TabPar} (column~4)
and correspond to the median point of the 21-cm line profile measured
at 20\% of maximum intensity. 

The internal mean error on $V_{20}$ is calculated according to
Fouqu\' e et al.\, (1990) as follows:
\[ \sigma(V_{20}) = \frac{4 \cdot (R \cdot \alpha)^{1/2}}{S/N} \]
where $R$ is the actual spectral resolution, $\alpha = 
(W_{20}-W_{50})/2$ is the slope of
the line profile, and $S/N$ is the signal to noise ratio.
The average of $\sigma (V_{20})$ is about 8 km.s$^{-1}$.

\paragraph{Line widths \label{LinWid}}

Line widths are measured on the observed profile at two standard
levels corresponding to 20\% and 50\% of the maximum intensity 
of the line. The results listed in Table \ref{TabPar}, columns 6 and 9, have been corrected to 
the optical velocity scale. We also provide line 
widths corrected for resolution effect (Fouqu\'e et al \cite{fou90}) in columns 7 and 10. 
The mean measurement error is taken equal to $3 \cdot \sigma(V_{20})$ and 
$2 \cdot \sigma(V_{20})$ for the 20\% and 50\% widths, respectively. 
The data presented here are not corrected for internal 
velocity dispersion. Details about these corrections can be found in 
Bottinelli et al (\cite{bot90}), Fouqu\'e et al (\cite{fou90}) or in Paturel et al (\cite{pat03b}).

\begin{center}
\begin{figure}
\epsfig{file=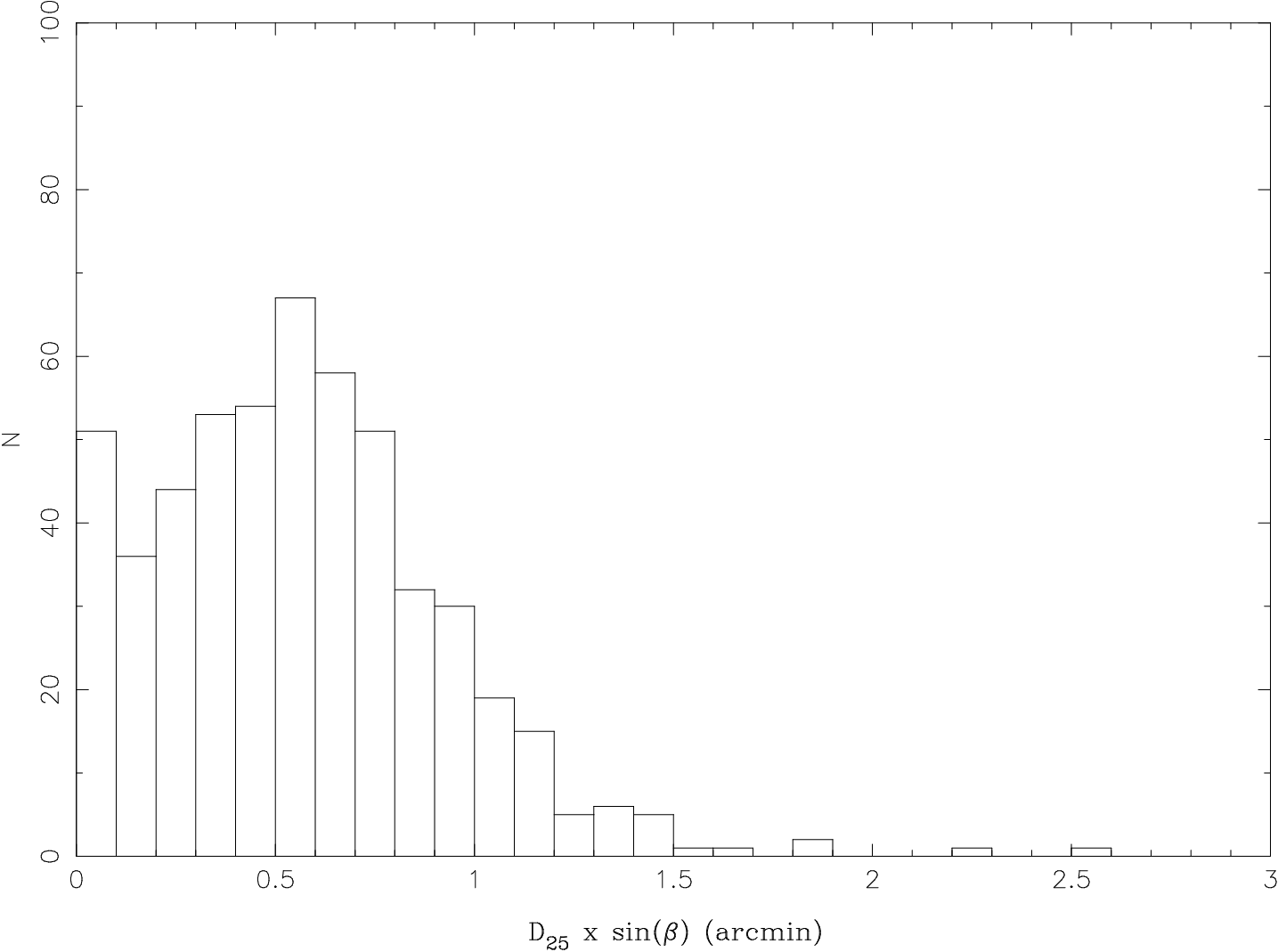,angle=270,width=8.5cm}
\caption[]
{Projection of the major axis $D_{25}$ on the East-West direction}
\end{figure}
\end{center}

\paragraph{HI-fluxes \label{HIflu}}

The detailed description of the flux calibration is given in Theureau et al. (\cite{the05}).

HI-fluxes $F_{HI}$ (Table \ref{TabPar}, column 12) are expressed in Jy km s$^{-1}$.
The values given in column 13 are corrected for beam-filling according to Paturel et al. (\cite{pat03b}):
\[ F_{HIc} = B_f \cdot F_{HI} \]
where $F_{HI}$ is the observed raw HI-flux, 
\[ B_f = \sqrt{(1+xT)(1+xt)} \]
\[ T= (a_{25}^2 \sin^2\beta + b_{25}^2 \cos^2\beta)/\theta_{EW}^2 \]
\[ t= (a_{25}^2 \cos^2\beta + b_{25}^2 \sin^2\beta)/\theta_{NS}^2 \]
$\theta_{EW}$ and $\theta_{NS}$ are the half-power beam dimensions
of the Nan\c cay antenna, $\beta$ is the position angle of the galaxy defined north-eastwards,
$a_{25}$ and $b_{25}$ are the photometric major and minor axis respectively. The parameter x 
is $x=0.72 \pm 0.06$ (Bottinelli et al., \cite{bot90}). The distribution of the East-West projection
of $D_{25}$ diameters is shown in Fig.\ 2. This is to be compared to the 4 arcmin width of the 
half-power beam.


\begin{table*}
\small{
\setlength{\tabcolsep}{0.03in}
\begin{tabular}{lllrrrrrrrrrrrrccc}
pgc/leda   &    NAME      & RA (2000) DEC &  $V_{20}$  &  $\sigma_V$  &  $W_{20}$ &  
$W_{20c}$ & $\sigma_{W20}$ &  $W_{50}$ & $W_{50c}$ & $\sigma_{W50}$ &  F(HI) & F(HI)$_{c}$ & $\sigma_F$ &  $S/N$ & rms & Q & flag \\
\hline

pgc0000115 & PGC000115    &  J000145.3-042049  &        &      &       &       &       &       &       &       &       &        &      &      & 2.2 & E &   \\
pgc0000287 & NGC7813      &  J000409.1-115902  &  9122. &  13. &  403. &  396. &   39. &  387. &  385. &   26. &  1.84 &  1.85  & 0.51 &  3.1 & 1.4 & B &   \\
pgc0000317 & ESO409-011   &  J000439.3-282738  &  8041. &   4. &  291. &  284. &   11. &  282. &  280. &    8. &  4.08 &  4.12  & 0.68 &  7.9 & 2.8 & A &   \\
pgc0000432 & ESO409-016   &  J000556.1-310610  &  7837. &   7. &  335. &  328. &   21. &  320. &  318. &   14. &  3.46 &  3.48  & 0.74 &  5.6 & 2.8 & A &   \\
pgc0001011 & NGC0054      &  J001507.7-070625  &  5333. &  18. &  429. &  422. &   55. &  413. &  411. &   37. &  3.17 &  3.23  & 1.22 &  2.2 & 3.2 & B &   \\
pgc0001165 & PGC001165    &  J001759.4-091620  &  6979. &  22. &  440. &  433. &   67. &  418. &  416. &   45. &  2.49 &  2.54  & 0.98 &  2.1 & 2.5 & C &   \\
pgc0001431 & PGC001431    &  J002221.4-012045  &  4898. &   6. &  173. &  166. &   17. &   77. &   75. &   11. &  7.47 &  7.52  & 0.75 & 17.2 & 4.1 & C & c \\
pgc0001453 & PGC001453    &  J002238.2-240733  &  9951. &  18. &  544. &  537. &   55. &  504. &  502. &   36. &  3.09 &  3.09  & 0.75 &  3.5 & 2.0 & C & c \\
pgc0001542 & NGC0102      &  J002436.5-135722  &  7333. &  26. &  464. &  457. &   77. &  438. &  436. &   51. &  1.64 &  1.66  & 0.67 &  2.0 & 1.7 & C &   \\
pgc0001813 & NGC0131      &  J002938.5-331535  &  1422. &  25. &  205. &  198. &   75. &  183. &  181. &   50. & 10.19 & 10.39  & 6.87 &  1.9 & 27.3 & C &   \\
pgc0001897 & PGC001897    &  J003058.2-092415  &        &      &       &       &       &       &       &       &       &        &      &      & 1.5 & E &   \\
pgc0002046 & ESO540-002   &  J003414.1-212812  &  6983. &   5. &  287. &  280. &   14. &  275. &  273. &    9. &  3.17 &  3.20  & 0.51 &  7.4 & 1.9 & A &   \\
pgc0002047 & ESO540-001   &  J003413.6-212619  &  8048. &   3. &  131. &  124. &    9. &   92. &   90. &    6. &  4.21 &  4.33  & 0.35 & 22.0 & 2.0 & A &   \\
pgc0002164 & PGC002164    &  J003611.3-323428  &  4381. &  16. &  202. &  195. &   48. &  162. &  160. &   32. &  1.93 &  1.95  & 0.63 &  4.0 & 2.6 & C & c \\
pgc0002333 & PGC002333    &  J003920.1-102855  & 10948. &  11. &  328. &  321. &   32. &  319. &  317. &   22. &  1.82 &  1.84  & 0.69 &  2.8 & 2.3 & B &   \\
pgc0002349 & NGC0178      &  J003908.5-141018  &  1447. &   3. &  138. &  131. &    8. &   99. &   97. &    5. &  9.82 &  9.99  & 0.74 & 22.9 & 4.0 & A &   \\
pgc0002352 & NGC0192      &  J003913.4+005151  &  4212. &  11. &  423. &  416. &   32. &  390. &  388. &   21. &  2.27 &  2.46  & 0.43 &  5.5 & 1.3 & C & c \\
pgc0002369 & ESO540-009   &  J003921.0-185503  &  3893. &  10. &  188. &  181. &   29. &  172. &  170. &   19. &  1.39 &  1.42  & 0.46 &  4.2 & 2.0 & B &   \\
pgc0002380 & NGC0187      &  J003930.2-143922  &  3937. &   9. &  319. &  312. &   28. &  293. &  291. &   19. &  4.29 &  4.33  & 0.81 &  5.5 & 2.6 & B &   \\
pgc0002424 & PGC002424    &  J004029.1-101820  &  8104. &  10. &  336. &  329. &   31. &  317. &  315. &   21. &  1.49 &  1.50  & 0.38 &  4.2 & 1.3 & B &   \\
pgc0002460 & UGC00435     &  J004059.6-013802  &  5438. &   5. &  299. &  292. &   15. &  281. &  279. &   10. &  2.57 &  2.62  & 0.47 &  8.8 & 2.3 & C & c \\
pgc0002476 & PGC002476    &  J004122.4-354836  &  6408. &   9. &  258. &  251. &   28. &  240. &  238. &   19. &  2.10 &  2.12  & 0.60 &  4.6 & 2.5 & B &   \\
pgc0002637 & IC1578       &  J004426.0-250434  &  6737. &  34. &  390. &  383. &  102. &  306. &  304. &   68. &  1.79 &  1.80  & 0.58 &  2.7 & 1.6 & C &   \\
pgc0002797 & ESO411-016   &  J004739.6-275651  &  1783. &   9. &  156. &  149. &   27. &  120. &  118. &   18. &  2.38 &  2.39  & 0.57 &  6.7 & 2.9 & A &   \\
pgc0002955 & ESO411-022   &  J005042.7-312302  &  5959. &  19. &  145. &  138. &   56. &   92. &   90. &   38. &  3.47 &  3.54  & 1.77 &  3.9 & 10.9 & C & ? \\
pgc0003636 & UGC00627     &  J010100.6+132806  & 11757. &      &  461. &  454. &       &  468. &  466. &       &  2.06 &  2.08  & 0.66 &  2.6 & 1.7 & C &   \\
pgc0003642 & ESO351-031   &  J010060.0-351433  &        &      &       &       &       &       &       &       &       &        &      &      & 2.3 & E &   \\
pgc0004316 & ESO352-018   &  J011209.4-321432  &  9893. &  11. &  246. &  239. &   34. &  237. &  235. &   22. &  1.43 &  1.47  & 0.64 &  2.7 & 2.4 & C & c \\
pgc0004352 & PGC004352    &  J011235.1-040824  &  5664. &  32. &  341. &  334. &   97. &  318. &  316. &   64. &  1.11 &  1.13  & 0.71 &  1.5 & 2.1 & D &   \\
pgc0004641 & ESO352-034   &  J011725.2-354702  &  9628. &  11. &  277. &  270. &   34. &  228. &  226. &   23. &  2.68 &  2.72  & 0.55 &  6.2 & 2.2 & B &   \\
pgc0004703 & ESO542-003   &  J011844.5-193736  &  6444. &  23. &  423. &  416. &   68. &  374. &  372. &   45. &  1.34 &  1.37  & 0.46 &  3.1 & 1.5 & C &   \\
pgc0005055 & UGC00928     &  J012312.4-003828  &        &      &       &       &       &       &       &       &       &        &      &      & 2.3 & E &   \\
    ...    &    ...       &    ...             &   ...  &  ... &   ... &   ... &   ... & ...   & ...   & ...   & ...   &  ...   & ...  & ...  & & ... & \\
\hline
\end{tabular}
}
\caption[]
{{\bf Astrophysical HI-parameters (excerpt)}.  \\
Column 1: PGC or LEDA galaxy name; \\
Column 2: most usual galaxy name; \\
Column 3:  J2000 equatorial coordinates; \\
Column 4: systemic heliocentric radial velocity (km s$^{-1}$); \\
Column 5: rms error (km s$^{-1}$); \\
Column 6: total line width at 20\% of the maximum intensity (km s$^{-1}$); \\
Column 7: total corrected line width at 20\% (km s$^{-1}$); \\
Column 8: rms error (km s$^{-1}$); \\
Column 9: total line width at 50\% of the maximum intensity (km s$^{-1}$); \\
Column 10: total corrected line width at 50\% (km s$^{-1}$); \\
Column 11: rms error (km s$^{-1}$); \\
Column 12: observed HI-flux (Jy km s$^{-1}$); \\
Column 13: beam-filling corrected HI-flux (Jy km s$^{-1}$); \\
Column 14: rms error (Jy km s$^{-1}$); \\
Column 15: signal to noise ratio; \\
Column 16: rms noise; \\ 
Column 17: quality code (see Sect. 2) \\
Column 18: flag ("c" indicates confirmed HI confusion with the emission of another galaxy; "?" means that confusion is suspected but not certain)}
\label{TabPar}
\end{table*}

\small{

\begin{table*}
\begin{tabular}{llrrll}
pgc/leda   & Type & $\log{D_{25}}$ & P.A. & Q & comments \\
 & & log(0.1') & deg. & & \\
\hline
pgc0000115 & Sab  &   0.87  &   92.5 & E  &   small galaxy group, also pgc 131 (MCG-01-01-023) close to the [...]  \\ 
pgc0000287 & Sb   &   0.91  &  158.0 & B  &   Theureau et al 2005                                                         \\
pgc0000317 & S0-a &   0.96  &   37.0 & A  &                                                                              \\
pgc0000432 & Sa   &   1.03  &    8.7 & A  &                                                                             \\
pgc0001011 & SBa  &   1.14  &   92.6 & B  &   Paturel et al 2003, pgc1024961 also in beam 3' NW, late type,  [...] \\
pgc0001165 & S0-a &   0.96  &  126.3 & C  &   SO galaxy                                                                 \\
pgc0001431 & Sbc  &   0.74  &   48.0 & Cc &   HI confusion w UGC212 at V=4840, multiple/interacting galaxy,  [...]     \\
pgc0001453 & ???  &   0.35  &        & Cc &   HI confusion w ESO473-018 at V=9923    Theureau et al 2005               \\
pgc0001542 & S0-a &   0.97  &  126.5 & C  &   SO-a B galaxy                                                             \\
pgc0001813 & SBb  &   1.25  &   62.5 & C  &   =NGC131, edge of NGC134 in cf at V=1579 but prob no HI confusion [...] \\
pgc0001897 & Sb   &   0.91  &  123.5 & E  &                                                                           \\   
pgc0002046 & SBb  &   0.98  &  125.4 & A  &   =ESO540-002, galaxy group, pgc2047 and pgc2057 in the beam at [...] \\
pgc0002047 & SBc  &   1.12  &  168.6 & A  &   pgc2057 (early type) also in the beam at V=8172                           \\       
pgc0002164 & S0-a &   0.95  &   40.8 & Cc &   HI confusion w ESO350-037=pgc2157 at V=4312                              \\   
pgc0002333 & Sab  &   0.91  &  174.0 & B  &                                                                             \\
pgc0002349 & SBm  &   1.33  &    3.0 & A  &   Bottinelli et al 1982, interacting or peculiar galaxy                     \\
pgc0002352 & SBa  &   1.31  &  164.9 & Cc &   HI confusion, galaxy group in the beam w NGC197,NGC196, [...]  \\
pgc0002369 & Sb   &   1.04  &  172.2 & B  &                                                                          \\
pgc0002380 & SBc  &   1.17  &  149.2 & B  &                                                                           \\
pgc0002424 & Sab  &   0.85  &   65.0 & B  &                                                                            \\
pgc0002460 & Sab  &   0.98  &   24.8 & Cc &   probably HI confusion w pgc090496 in cf, V unknown                     \\
pgc0002476 & SBbc &   0.95  &   56.0 & B  &                                                                           \\
pgc0002637 & Sb   &   0.93  &   17.7 & C  &   Theureau et al 2005                                                      \\
pgc0002797 & Sab  &   1.01  &   46.5 & A  &                                                                            \\
pgc0002955 & SBb  &   1.10  &  168.5 & C? &   ESO411-021 Irr in beam 3' N, unknown V, interaction ? probably [...] \\
pgc0003636 & Sa   &   0.88  &  118.5 & C  &                                                                             \\
pgc0003642 & SBb  &   0.93  &  174.3 & E  &   successfully observed by Theureau et al 2005                              \\
pgc0004316 & Sa   &   1.04  &  128.1 & Cc &   confusion with NGC0427 (Sa, V=10012) at the edge of the beam 11' [...] \\
pgc0004352 & Sa   &   1.05  &   89.0 & D  &   lenticular ? low SNR, pgc4346 E galaxy V=5611 also in beam                \\
pgc0004641 & SBb  &   0.98  &  171.6 & B  &   multiple object ? HI spectrum OK                                          \\
pgc0004703 & S0-a &   1.00  &   62.4 & C  &   low SNR                                                                  \\
pgc0005055 & S0   &   1.01  &   47.7 & E  &   t=-1.8                                                                  \\
pgc0005112 & S0-a &   0.98  &  164.4 & E  &   t=-0.5 barred galaxy                                                   \\
pgc0005145 & S0-a &   1.02  &   64.0 & E  &   t=0.0                                                                 \\
pgc0005151 & Sb   &   1.05  &   30.8 & D  &   Theureau et al 2005                                                   \\
pgc0005168 & SBb  &   0.76  &  107.7 & C  &   Theureau et al 2005                                                  \\
pgc0005505 & Sa   &   0.94  &   72.0 & B  &   pgc737697 V=9232 at edge of beam 3' SE                              \\
 ...  & ...  & ... \\
\hline 
\end{tabular}
\caption[]
{{\bf Notes on HI-observations (excerpt).}  \\
Column 1: PGC or LEDA galaxy name \\
Column 2: morphological type from HYPERLEDA\\
Column 3: logarithm of isophotal D$_{25}$ diameter in 0.1 arcmin from HYPERLEDA\\
Column 4: Major axis position angle (North Eastwards) from HYPERLEDA\\
Column 5: quality code and HI-confusion flag "c" (confirmed) or "?" (possible) (see Sect. \ref{SecSam}) \\
Column 6: comments; conf="HI confusion", comp="companion", cf="comparison field", poss="possible",w="with" }
\label{TabCom}
\end{table*}
}

\begin{figure*}
\epsfig{file=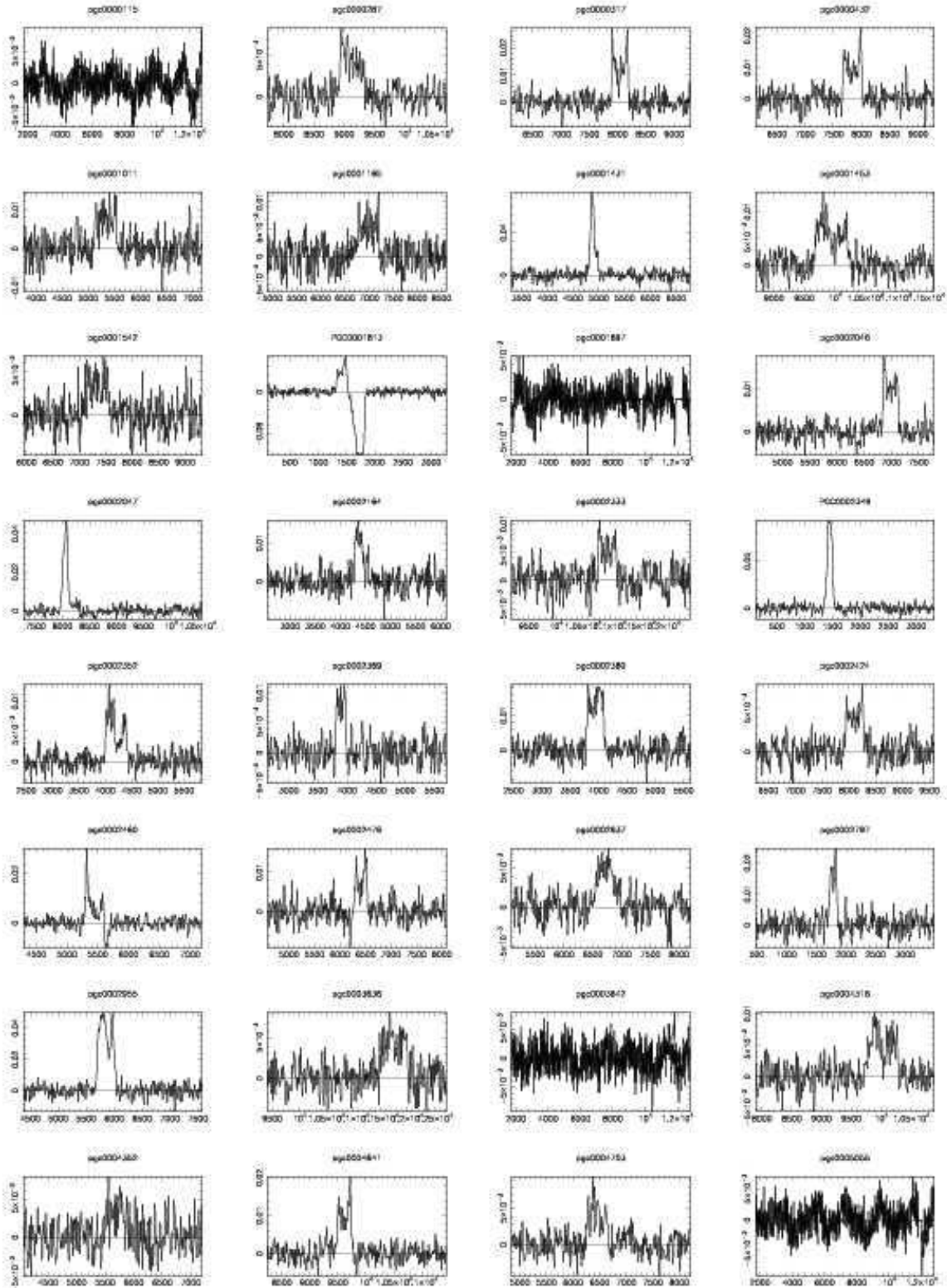,width=17cm}
\caption[]  
{{\bf -excerpt. }21-cm line profiles of galaxies listed in Table \ref{TabPar}; profiles 
are classified according to their PGC name which 
is written above each panel. Ordinate and abscissae axes are 
graduated respectively in km s$^{-1}$ and Jy. Note that heliocentric
radial velocities are expressed in terms of optical redshift 
$c \frac{\Delta \lambda}{\lambda}$. The horizontal line
represents the baseline of the profile, i.e. the zero flux level, from which the maximum
is estimated.} 
\end{figure*}


\section{Building the Tully-Fisher sample \label{SecSam}}

\subsection{The input data}

\paragraph{Rotational velocities}
                                                                                                                      
Rotational velocities, i.e. the $\log{V_m}$ parameter used in the Tully-Fisher relation,
have been mainly gathered from the HYPERLEDA compilation (16\,666 galaxies,
Paturel et al. \cite{pat03a}) and were complemented by some of our own recent HI line measurements
with the Nan\c cay radiotelescope (586 late type galaxies, Theureau et al. \cite{the05}).
A few other measurements from Haynes et al. (\cite{hay99}) not previously in HYPERLEDA were also added.
All our own 3300 HI spectra acquired with the Nan\c cay radiotelescope antenna in the last decade
were reviewed and assigned a quality code according to the shape of their 21-cm line profile.
This study (Guilliard et al. \cite{gui04}) allows to flag efficiently several TF
outliers due to morphological type mismatch or HI-confusion in the ellongated beam of Nan\c cay.
Even if this Nan\c cay subsample concerns only a part of the data ($\sim$ 25\%), we substantially
improved the apparent scatter of the TF relation.

In the HYPERLEDA compilation, the $\log{V_m}$ parameter is calculated from 21-cm line widths at 
different levels and/or rotation curves (generally in H$_{\alpha}$). 
The last compilation provides us with 50520 measurements of 21-cm line widths or maximum rotation velocity. 
These data are characterized by some secondary parameters: telescope, velocity resolution, level of the 21-cm 
line width. For data homogenization, HYPERELEDA uses the so-called EPIDEMIC METHOD (Paturel, G. et al. \cite{pat03b}).
One starts from a standard sample (a set of measurements giving a large and homogeneous sample : here, 
the Mathewson et al. \cite{mat96} data), all other 
measurements are grouped into homogeneous classes (for instance, the class of measurements made 
at a given level and obtained with a given resolution). The most populated class is cross-identified with 
the standard sample in order to establish the equation of conversion to the standard system. 
Then, the whole class is incorporated into the standard sample. So, the standard sample is growing progressively. 
The conversion to the standard propagates like an epidemy.
In summary, this kind of analysis consists in converting directly the widths for a given resolution r and given 
level l into a quantity which is homogeneous to twice the maximum rotation velocity ($=\log{W}$, uncorrected 
for inclination. A final correction is applied reference by reference to improve the homogenization.

The final $\log{V_m}$ value is corrected for inclination : $ logV_m = (\log{W} - \log{(2\sin{incl})} $.
Where the inclination $incl$ is derived following RC3 (de Vaucouleurs \cite{dev91}):
\[ sin^2(incl)= \frac{1-10^{2logR_{25}}}{1-10^{2logr_0}} \]
$R_{25}$ is the axis ratio in $B$ at the isophot corresponding to 25 mag\.arcsec$^{-2}$, $\log{r_0}$= 0.43 + 0.053$T$
for type $T$= 1 to 7 ($Sa$--$Sd$) and $logr_0$=0.38 for $T$=8 ($Sdm$).

\paragraph{Magnitudes}

The 2MASS survey, carried out in the
three infrared bands J, H and K, collected photometric data for 1.65 million galaxies
with $K_s<14$ (Jarrett et al. \cite{jar00}) and made the final extended source catalog
recently public.
Total magnitude uncertainties for the 2MASS extended objects are generally
better than 0.15 mag.
We exclude any galaxy with the accuracy of magnitudes worse than 0.3.
This accuracy appears reasonable when considering that it is almost impossible to obtain total
magnitudes better than 0.1, due to the difficulty to extrapolate the profile in a reliable way.

\paragraph{Extinction}

The extinction correction we applied includes a Galactic component, $a_{\rm G}$,
adopted from Schlegel et al.\ (\cite{sch98}),
and a part due to the internal absorption of the observed
galaxy, $a_{\rm i}$. Both depend on the wavelength.
\begin{equation}
a_{\rm ext} = f_{\rm G}(\lambda) a_{\rm G} +f_{\rm i}(\lambda) a_{\rm i}.
\end{equation}
The Galactic and internal wavelength conversion factors are
$f_{\rm G}(\lambda)=$1.0, 0.45, 0.21, 0.13, 0.085 (Schlegel et al. \cite{sch98})
and $f_{\rm i}(\lambda)=$ 1.0, 0.59, 0.47, 0.30, 0.15 (Tully et al. \cite{tul98}, Watanabe et
al. \cite{wat01}, Masters et al. \cite{mas03}) for B, I, J, H, and K bands, respectively.

\paragraph{Radial velocities}
                                                                                                                      
All radial velocities were taken from HYPERLEDA,  often a mix
and an average of several publications and redshift surveys. Within the limit of 8\,000 km~s$^{-1}$,
we collected a sample of 32\,545 galaxies. In general, when an HI measurement was available (i.e.
for most of the TF subsample used in this study), the radio radial velocity was prefered, being
more acurate than the available optical one. As it is the use in HYPERLEDA, when more than two
redshift measurements for the same galaxy were available, the most discrepant ones were rejected from
the mean. The radial velocities are used at the first iteration of the IND method
as a reference distance scale for the Malmquist bias correction.

\subsection{Selection and completeness}

The analytical treatment of the Malmquist bias effect with distance, by applying
the Iterative Normalized Distance method (IND), requires the strict completeness
of the samples according to magnitude selection (Theureau et al. \cite{the98b} and Sect. 4.).
The limits in magnitude are simply determined by eye as the knee observed in a $\log{N}$ vs.\ magnitude diagram,
witnessing the departure from a homogeneous distribution in space with growing distance.
This limit is in 'observed apparent magnitude', independently of extinction or opacity correction. These
corrections however are taken into account further, as part of the ND scheme itself, in what we call the effective
magnitude limit (Sect. 4.1.).
                                                                                                                                       
The adopted completeness limit are the followings :
J$_{\rm lim}$=12.0, H$_{\rm lim}$=11.5, and K$_{\rm lim}$=11.0 (equivalent
to B$_{\rm lim} \sim$ 15 and I$_{\rm lim} \sim$ 13).
Only the complete part of the sample in each band, about half of
available data, is included for further study.
                                                                                                          
The final selection is made according to the following conditions:
\begin{itemize}
\item J, H, and K magnitudes $<$ completeness limit
\item magnitude uncertainty $<$ 0.3 mag
\item log of rotational velocity uncertainty $<$ 0.03
\item $T$ = 1--8 to keep only fair spiral galaxies
\item $\log R_{25} > 0.07$ to avoid face-on galaxies for which the rotational velocity is poorly determined
                                                                                                          
\end{itemize}
                                                                                                          
\begin{figure}
\includegraphics[angle=270,width=8cm]{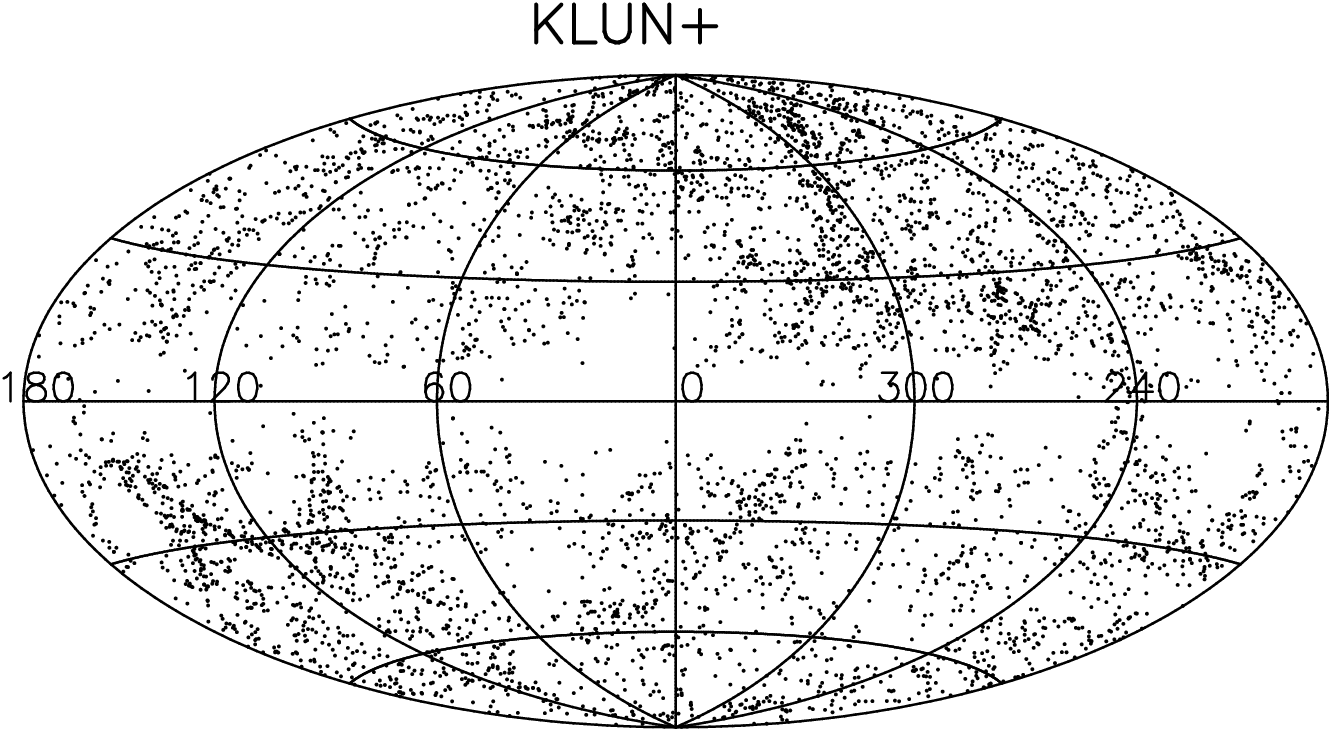}
\caption{Sky distribution of KLUN galaxies used in the current analysis}
\label{FigSky}
\end{figure}

After these restrictions there are 3263 spiral galaxies 
distributed over the whole sky (see Fig. \ref{FigSky}).

\section{Method of analysis \label{SecND}}

\begin{figure*}
\includegraphics[width=16cm]{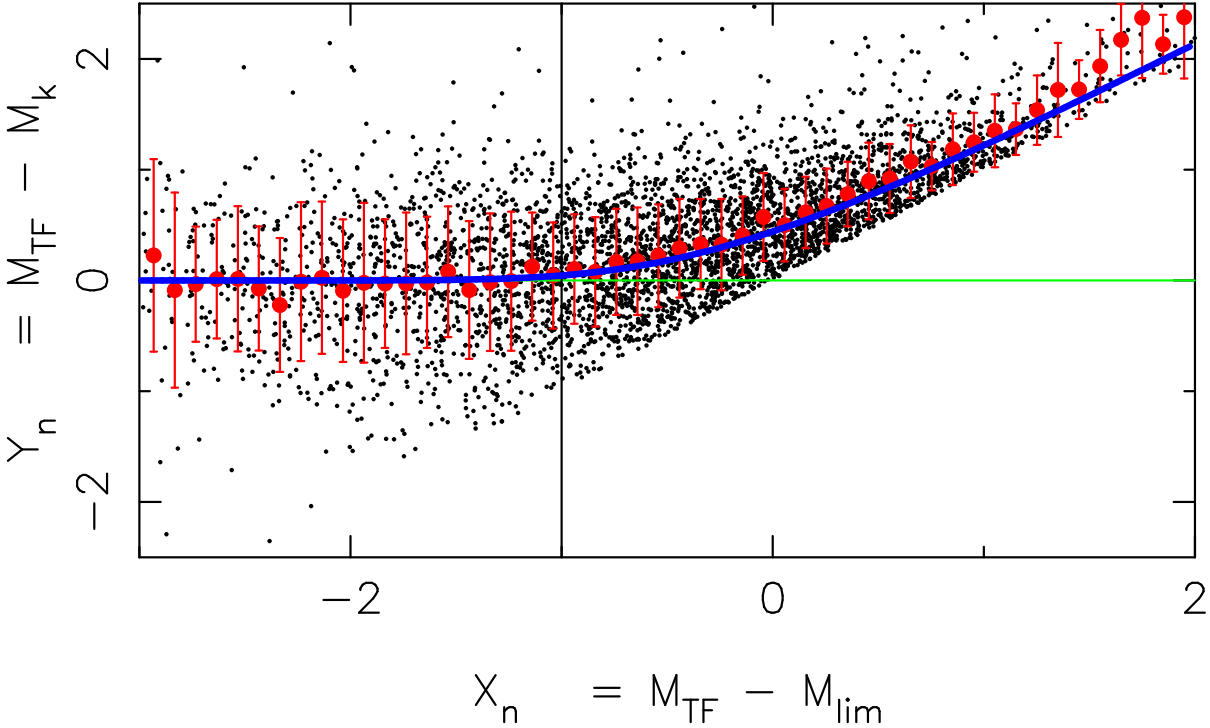}
\caption{TF residuals ($Y_{\rm n}$ against the normalized distance modulus ($X_{\rm n}$).
The ``unbiased plateau'' is the region at $X_{\rm n}<-1$, where the binned average
of the residuals, the balls with error bars, are close to the zero line. The
curve going through the balls is the analytical correction formula described by Eq. 8 in Sect.\
\ref{SecNor}.}
\label{FigRes}
\end{figure*}

In this section we explain the Iterative Normalized Distance method
for deriving the peculiar velocities.
The 'iterative' means that a previously calculated peculiar velocity field
is used for a more accurate estimation of new peculiar velocities.
The 'normalized distance' is a quantity depending on the distance and the
absolute size or absolute magnitude of a galaxy, such that for any galaxy, the average
selection bias (in the terminology of Strauss \& Willick, \cite{str95}) or the Malmquist bias
of the second kind (according to Teerikorpi, \cite{tee97}) can be given
by a function depending on its normalized distance,
the dispersion of the distance criterion, and the completeness limit.
This is illustrated in Fig.\ \ref{FigRes}, where the TF residuals, plotted
against the normalized distance modulus, clearly show the unbiased regime and
the deviation due to the magnitude cutoff.

A fully detailed description of the method follows, but we start by listing
the main steps:
\begin{enumerate}
\item Calculate the absolute magnitudes and the normalized distances
using the kinematical (redshift) distances.
\item Calculate the TF relation
using the unbiased part of the normalized distance diagram.
\item Use the unbiased TF relations and the analytical Malmquist correction
formula for estimating real space galaxy distances beyond the unbiased plateau limit (Fig\. 5).
\item Obtain the peculiar velocity field in a Cartesian grid in the redshift
space by smoothing the individual peculiar velocities given by the
Malmquist corrected TF distances.
\item Go back to step 1, and use the corrected kinematical distances by
substracting the smoothed peculiar velocity field values from the redshift
velocities.
\end{enumerate}
This loop is repeated until converging values for the peculiar velocities
are obtained. The peculiar velocities for all the galaxies do not converge
nicely, though. We thus extract the most unreliable galaxies (about 4 \%)
and recalculate the velocities with the reduced data set (see Sect. 4.4. and Fig. 6).
As confirmed by the tests done with a mock sample in Sect. 4.6. outliers use to
be mainly very low Galactitic latitude objects for which the corrected total magnitude is not
well estimated from observed one, and core cluster members whose observed radial velocity, used
as kinematical distance at the first iteration, does 
not reflect at all their true distance, due to their strong motion in the cluster potential.

\subsection{Kinematical distances, normalized distances and absolute magnitudes \label{SecNor}}
                                                                                                                      
Let us define the kinematical distance modulus as
\begin{equation}
\mu_{\rm k} = 5 \log\frac{cz}{H_0} +25,
\label{EqMuk}
\end{equation}
where $cz$ is the observed heliocentric redshift, corrected by the CMB dipole motion. The absolute magnitude is
\begin{equation}
M_{\rm k}=m_0^{\rm c}-\mu_{\rm k}
\end{equation}
where $m_0^{\rm c}$ is the apparent magnitude, corrected for inclination,
extinction, and cosmological effects, as stated in Eq.\ 10 of
Paturel et al.\ (\cite{pat97}) (the cosmological correction is negligible for
all galaxies in the present study).
                                                                                                                      
If we consider that the TF relation is a linear law characterized
by a given slope and a given dispersion (the zero-point being fixed either
by some local calibrators or by adopting a value of $H_0$), and if we assume
that the sample is actually complete up to a well defined apparent magnitude limit,
then the selection bias at a fixed $\log{V_{\rm m}}$ is only a function of the distance
(see Teerikorpi \cite{tee84}, Theureau et al. \cite{the97}, \cite{the98b}).
In other words, the bias at a fixed $\log{V_{\rm m}}$ and at a given distance is only
the consequence of the magnitude cut-off in the distribution of TF residuals
and moreover it does not depend at all on space density law.
                                                                                                                      
By normalizing to a same luminosity class, i.e.\ a same $\log{V_{\rm m}}$
value, and by taking into account the variation of the actual magnitude cut-off with extinction,
one can build a unique diagram showing the bias evolution with distance.
                                                                                                                      
The distances and magnitudes are then scaled so that a sharp edge is seen at the sample
completeness limit.
                                                                                                                      
The normalized distance modulus is defined as
\begin{equation}
X_{\rm n} =  \mu_{\rm k} + (M_{\rm TF} - m_{\rm lim} + a_{\rm ext})
\end{equation}
where $M_{\rm TF}$ is the absolute magnitude of a galaxy, as given by the
Tully-Fisher relation and $m_{\rm lim}$
is the apparent magnitude completeness limit.
More explicitely, if we develop $M_{\rm TF}$ as $[a_1 \log{V_{\rm m}} + a_0]$ it appears
that we normalize indeed to the same $\log{V_{\rm m}}$ and the same effective magnitude
limit $m_{\rm lim}^{\rm eff} = m_{\rm lim} - a_{\rm ext}$.
$X_n$ can also be expressed as $M_{\rm TF} - M_{\rm lim}$, i.e. the difference
between the TF absolute magnitude at a given $\log{V_{\rm m}}$ and the absolute magnitude
cut off (in the TF residuals) at a given distance.
                                                                                                                      
The normalized magnitude
\begin{equation}
Y_{\rm n} = M_{\rm TF}- M_{\rm k} = M_{\rm TF} - m_0^{\rm c} + \mu_{\rm k}
\end{equation}
corresponds to the departure of the absolute magnitudes calculated on the basis of
kinematical distances from the true mean value given by the TF relation. This residual
contains the contribution of magnitude and $\log{V_{\rm m}}$  measurement
errors, internal TF dispersion, and peculiar velocity.
                                                                                                                      
Figure \ref{FigRes} shows normalized distance moduli vs. normalized magnitudes
for the galaxies derived by the Tully-Fisher relation.
The curve going through the points is the analytical bias solution
$f(X_{\rm n},\sigma_{\rm TF},m_{\rm lim})$ while the vertical line shows the upper limit
adopted for the unbiased normalized distance domain.

\subsection{Unbiased Tully-Fisher}
                                                                                                                      
The TF relation states the linear connection between absolute magnitude
and $\log V_{\rm m}$ 
\begin{equation}
M=a_1 \log V_{\rm m} + a_0
\end{equation}
                                                                                                                      
One gets the slopes, $a_1$, and zero-points, $a_0$, of the TF relations in each band ($IJHK$) by least
squares fit using the unbiased part of each sample. The unbiased part is the
flat or plateau region in the normalized distance diagram. This subsample also provides
us with the "0-value" of the TF residuals that is used to compute the bias deviation for
the whole magnitude complete sample (the horizontal line in Fig. \ref{FigRes}).
The method of deriving the TF relation in the unbiased plateau has been succesfully used in previous
KLUN studies; a full statistical description can be found in Theureau et al. \cite{the98b} and
it was even numerically tested by Ekholm (\cite{ekh97}).

In the iterative scheme one starts by assuming a priori values for the TF slope and zero-point
(here, only the slope is important and a rough value can be inferred directly from the whole sample).
These values are used to compute the normalized distance and extract a first unbiased subsample.
The loop "TF-slope $\rightarrow$ normalized-distance $\rightarrow$ unbiased-subsample
 $\rightarrow$ TF-slope" can be repated a couple of times to be sure to start on the basis of
unbiased values.
                                                                                                                      
\subsection{Corrected distances}
                                                                                                                      
The function $f(X_{\rm n},\sigma_{TF},m_{\rm lim})$ is deduced from the expectancy of $Y_{\rm n}$
knowing $X_{\rm n}$ assuming that the magnitude selection $S(m)$ is described by the Heavyside function
$S(m)=\theta(m-m_{\rm lim})$. We have then:
\begin{equation}
E(Y_{\rm n}|X_{\rm n}) = \frac{1}{B} \int_{Y_{\rm lim}}^{-\infty} \frac{1}{\sqrt{2\pi} \sigma} Y \exp[-\frac{Y^2}{2 \sigma^2}] dY
\end{equation}
where
\[ \sigma = \sigma_{\rm TF} \]
\begin{eqnarray*} Y_{\rm lim}& =& Y_{\rm n}(m_{\rm lim}) = M_{\rm TF} - m_{\rm lim}^{\rm eff} + \mu_{\rm k} \\
&=& M_{\rm TF} - (m_{\rm lim} - a_{\rm ext}) + \mu_{\rm k}\\
&\equiv& X_{\rm n}
\end{eqnarray*}
\[ B = \int_{Y_{\rm lim}}^{-\infty} \frac{1}{\sqrt{2\pi} \sigma} \exp[-\frac{Y^2}{2 \sigma^2}] dY \]
From here we derive:
\begin{equation}
f(X_{\rm n},\sigma,m_{\rm lim}) = \sigma \, \sqrt{\frac{2}{\pi}} \, \frac{e^{-A^2}}{{\rm erfc}(A)}
\end{equation}
where,
\[ A = - \frac{1}{\sqrt{2} \sigma} X_{\rm n} \]
and
\[ {\rm erfc}(x) = \frac{2}{\sqrt{\pi}} \int_{x}^{-\infty} {e^{-y^2}dy} \]
The corrected and unbiased distance modulus is then finally:
\[ \mu_c = m^c - M_{\rm TF} + f(X_{\rm n},\sigma,m_{\rm lim}) \]
                                                                                                                      
Note that $H_0$ is cancelled out in $Y_{lim}$: it is indeed hidden
in the TF zero-point and explicitely present in $\mu_k$ but with an opposite sign.
                                                                                                                      
The reader will remark that our approach to the bias in this paper is radically different
to what has been attempted with MarkIII (Willick et al \cite{wil97} or Dekel et al \cite{dek99})
in which the approach has been from the viewpoint of the classical Malmquist bias
(using in particular some inhomogeneous density correction).
                                                                                                                      
\subsection{Iterations}
                                                                                                                      
The peculiar velocities of galaxies are then smoothed onto a cartesian grid,
$v_{\rm pec}(x,y,z)$ (Sect. \ref{SecSmo}).
As explained above, the method relies on kinematical
distances; the normalized distances and the absolute magnitudes needed
for TF relation include the kinematical distance. These distances can
be made more accurate by subtracting the peculiar velocities from
the redshift:
\begin{equation}
\mu_{\rm k} = 5 \log\frac{cz-v_{\rm pec}(x,y,z)}{H_0} +25.
\label{EqMuKC}
\end{equation}
This distance modulus then replaces the value given by
Eq.\ \ref{EqMuk}, and we can repeat the whole procedure with these updated distances.
The new peculiar velocity field is again used for correcting the kinematical
distances in the next iteration, and we keep on repeating the process until
convergence.

\begin{figure}
\includegraphics[width=8cm]{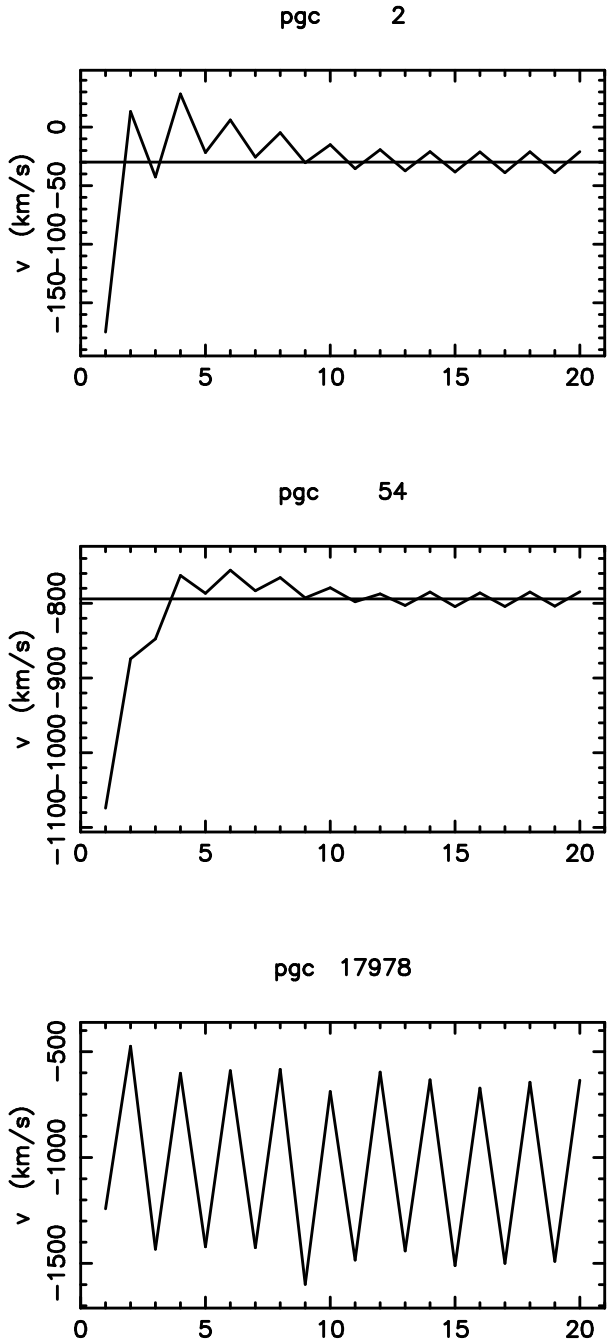}
\caption{The evolution of peculiar velocities with respect to the number
of iterative loop. Here are the two first galaxies in our sample,
and an exceptionally ``bad'' case (pgc 17978) which was rejected from the
final analysis.}
\label{FigVpEv}
\end{figure}

In practise, subtracting the whole $v_{\rm pec}(x,y,z)$ as in Eq.\ \ref{EqMuKC} would
overcorrect for the peculiar velocities, and cause diverging oscillatory
behaviour in the iterative process. Using a scaling factor
$\lambda \in (0.0,1.0)$ so that
\begin{equation}
\mu_{\rm k}^{j+1} = 5 \log\frac{cz-\lambda\,v_{\rm pec}^j(x,y,z)}{H_0} +25,
\end{equation}
removes this problem. The superscript $j$ corresponds to the iteration
number. Using the value $\lambda=0.5$ we reach converging values
after about 5--10 iterations. Figure \ref{FigVpEv} shows this convergence
for a few galaxies. Usually the $v_{\rm pec}^j$ approaches nicely to
a constant value, in some cases the values oscillate even with high $j$.
We checked all these convergence curves by eye and rejected the
worst cases. With the restricted sample we recalculated the peculiar velocities
and used the results of the selected 3126 spirals
galaxies for the peculiar velocity mapping.

\subsection{The tensor smoothing \label{SecSmo}}

After deriving radial peculiar velocities of galaxies it is useful
to interpolate these velocities at uniformly distributed grid points.
The best method is to smooth the observed galaxy velocities
with an appropriate window function. Dekel et al. (\cite{dek99}) discuss
the problems of smoothing a non-uniformly distributed set of radial velocities:
                                                                                                            
The radial velocity vectors are not all pointing in the same
direction over the smoothing window. Then, for example in a case of a
pure spherical infall towards the window center, all the transverse
velocities are observed as negative radial velocities. The net velocity
in the smoothing window is then, incorrectly, negative, in stead of being zero.
Dekel et al. (\cite{dek99}) call this the tensor window bias. They find
that it can be reduced by introducing a local velocity field with extra
parameters, which is to be fitted for the observed radial velocities
in the smoothing window. The best results are obtained by constructing
a three-dimensional velocity field with a shear,
\begin{equation}
{\bf v}({\bf x}) = {\bf B} + {\bf L} \cdot ({\bf x} - {\bf x}_{\rm c}),
\end{equation}
where ${\bf L}$ is a symmetric tensor, and ${\bf x}_{\rm c}$ is
the window center. There are then nine free parameters, three for the
actual window center velocity ${\bf B}$ and the six components of
the tensor ${\bf L}$.
The resulting grid point velocity is just ${\bf v}({\bf x}_{\rm c}) = {\bf B}$.
                                                                                                                      
Furthermore, if the true velocity field has gradients within the effective
smoothing window, a nonuniform sampling will cause an error, called the
sampling-gradient bias. Dekel et al.\ (\cite{dek99}) suggest that this
bias can be diminished by weighting the observed galaxy velocities by the
volume $V_n$, which is defined as the cube of the distance between the
galaxy and its $n$th neighbour. This method gives more weight for galaxies
in isolated areas.
                                                                                                                      
\subsection{Testing the method}
We test these biases with a mock peculiar velocity catalog.
The mock catalog is constructed from the GIF consortium constrained n-body
simulation of our $80 h^{-1}$~Mpc neighbourhood (Mathis et al.\
\cite{mat02}, http://www.mpa-garching.mpg.de/NumCos/CR/).
The simulation was run for a flat $\Lambda$CDM cosmological model, and it
provides locations, velocities, masses, and luminosities, with and without the
internal absorption, of 189\,122 galaxies.
The galaxy formation was defined by applying
a semianalytic algorithm on the dark matter merger tree. We added the Galactic
component of the absorption, as defined in Schlegel et al.\ (\cite{sch98}),
and selected the galaxies brighter than a magnitude limit. In the end there are 
9800 galaxies with their apparent B band magnitude smaller than 14.5.

\begin{figure*}
\includegraphics[width=16cm]{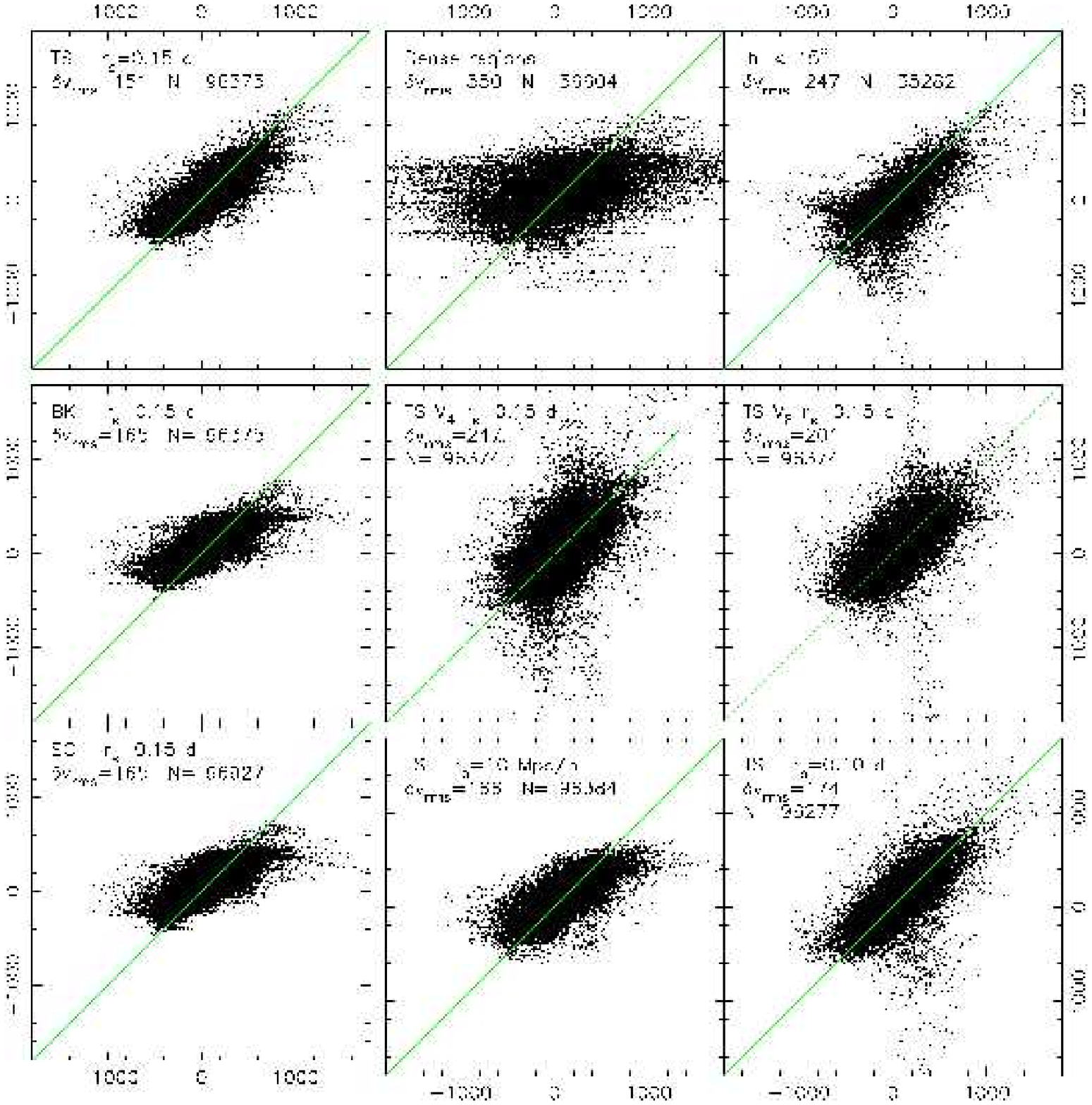}
\caption{Shown here are the true peculiar velocities (on y axis)
vs.\ the velocities obtained with a smoothing method for the GIF
consortium simulated data. The upper left corner of each panel gives the
essential information about the plot; TS stands for the tensorial smoothing,
using nine parameters, the bulk velocity {\bf B} and the tensor {\bf L},
BK is the three parameter bulk smoothing (only {\bf B}), and SC means
scalar smoothing, using a simple Gaussian smoothing window. $r_s$ is the
smoothing radius, either a distance dependent value, e.g. $0.15 d$ or
a fixed value, e.g. 10~Mpc~$h^{-1}$. The rms difference between the smoothed and the true
velocity field, $\delta v_{\rm rms}$, is given in km~s$^{-1}$. N is the number of points,
the simulated galaxies, in each figure. Upper left figure shows the
smoothing giving the smallest $\delta v_{\rm rms}$. In each figure the
galaxies in densest regions were excluded, as well as the objects at
the low Galactic latitudes. The two panels in the upper right corner
show how these points give substantially more divergent values.
The ``Dense regions'' and ``$|b|<15^o$'' maps were obtained with
the TS $r_s=0.15 d$ smoothing. V$_4$ and V$_8$ correspond to the
volume weighted method, using the volume defined by the fourth and the
eight closest galaxy, respectively. The volume weighting is claimed to
reduce the sampling gradient bias, but in this test they fail to produce
a better estimate of the true velocity field than the non-weighted method.
}
\label{FigSmo}
\end{figure*}

Figure \ref{FigSmo} shows the true vs. smoothed peculiar velocities
using different smoothing method and set of window parameters. These plots lead to the
following comments:
\begin{itemize}
\item the scatter is mainly related to the smoothing radius
\item outliers are essentially: \\
- galaxies at low Galactic latitude, for which the magnitude is not well defined and where
the sparser sampling leads the smoothing to diverge (vertical spreading) \\
- galaxies belonging to clusters for which the kinematical distance derived from the redshift
is strongly affected by the cluster velocity dispersion
\item one observes a tilt with respect to the line with slope = 1, when no tensor window is used.
This is the effect of the velocity gradient around structures and leads to an underestimation
of infall patterns and to a cooler velocity field.
\end{itemize}
                                                                                                          
The best $\chi^2$ value is obtained for $\rm B+L$ smoothing, and $R_{smooth}=0.15 \times distance$.
This will be our choice of smoothing for the rest of the study.

\section{Preliminary results \label{SecRes}}

In this section we present the TF relation parameters obtained for three wavelengths (JHK)
and some examples of maps of radial peculiar velocity fields, superimposed on the 
distribution of galaxies. A more detailed kinematical study is beyond the scope of the 
present analysis and will be presented in a forthcoming paper.

\subsection{TF parameters}
                                                                                                                      
\begin{figure}
\includegraphics[scale=0.7,angle=270]{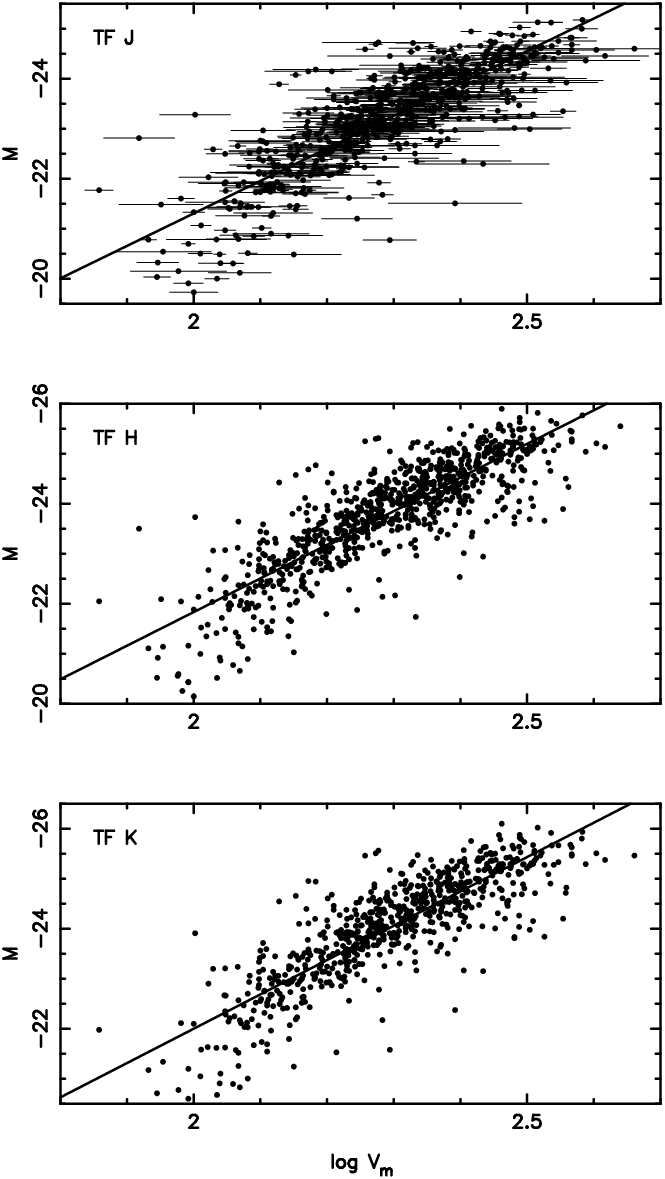}
\caption{Tully-Fisher relations in J, H, and K bands for unbiased
plateau galaxies. See Table \ref{TabTF} for the parameters of the relations.}
\label{FigTF}
\end{figure}

Figures \ref{FigTF} show the final TF relations for the galaxies in the
unbiased part of the normalized distance diagram. Tables \ref{TabTF}-- list the
parameters $a_1$ and $a_0$ and the scatter of the relation :
\begin{equation}
M=a_1 \log V_{\rm m} + a_0
\end{equation}

\begin{table}
\caption{Tully-Fisher parameters: slope, zeropoint, scatter, and the number
of unbiased plateau galaxies used in the relation.}
\begin{tabular}{ccccr}
\hline
 & $a_1$ & $a_0$ & $\sigma_{\mathrm TF}$ & $N$ \\
\hline
J & -6.3 $\pm$ 0.31 & -8.8  $\pm$ 0.14 & 0.46 & 960 \\
H & -6.4  $\pm$ 0.33 & -9.1  $\pm$ 0.14 & 0.47 & 1166 \\
K & -6.6  $\pm$ 0.37 & -9.0  $\pm$ 0.16 & 0.45 & 990 \\
\hline
\end{tabular}
\label{TabTF}
\end{table}
                                                                                                                      
The observed scatter is comparable to what was found by Karachentsev et al \cite{kar02} using 2MASS magnitudes.
The small difference (we get slightly smaller $\sigma$'s) can be explained easily by
our optimization of the kinenatical distance scale through the iterative process described
above. Accounting for the observed broadening due to apparent magnitude and $\log{V_{max}}$
uncertainties and to the residual peculiar velocity dispersion affecting the kinematical distances,
one obtains an internal scatter of $\sim$ 0.44 mag for the TF relation in $B$ and $\sim$ 0.4 mag in $K$.
This is 0.1 mag  greater than in studies restricted to pure rotation curve measurements of $\log{V_{max}}$.
Here instead, a large majority of $\log{V_{max}}$ measurement are from the width of global HI profiles.
As we know, even once corrected for non-circular motions, this width is still determined by the shape of
a galaxy's rotation curve, the distribution of HI gas in the disk and the possible presence of a warp
(Verheijen \cite{ver01}), leading to a greater intrinsic Tully-Fisher scatter.

\subsection{Peculiar velocities}

Table \ref{TabDis} shows the first 10 KLUN+ galaxies with TF distances. The distances
are expressed in km~s$^{-1}$. That is followed by the kinematical distance,
corrected by the radial peculiar velocity. This is our estimate of the
true distance of a galaxy. Finally we list the observed redshift velocity
in the CMB rest frame. The full catalog is
given in electronic format only.
\begin{table*}
\caption{An excerpt of the  table showing the distance data. We list the
name and the galactic coordinates of the galaxies, followed by the three
TF distances. Then we give the corrected kinematical
distance, $d_{\rm k}^{\rm c}=V_{\rm 3k}-V_{\rm pec}$. All are expressed in km~s$^{-1}$. The last entry is the redshift velocity,
corrected to the CMB rest frame. }
\begin{tabular}{lrrrrrrr}
\hline
&$l$&$b$& $d_{\rm TF,J}$ & $d_{\rm TF,H}$ & $d_{\rm TF,K}$  &$d_{\rm k}^{\rm c}$ & $V_{\rm 3k}$ \\
\hline
PGC 2 &   113.96 &   -14.70 &   6084 &   6030 &   5992 &   4762 &   4751 \\
PGC 54 &   109.57 &   -33.17 &   8864 &   8936 &   9152 &   8771 &   8379 \\
PGC 76 &   109.81 &   -32.67 &   7997 &   7592 &     ---  &   6968 &   6570 \\
PGC 102 &   111.34 &   -27.22 &   5214 &   5125 &   5211 &   4834 &   4720 \\
PGC 112 &   110.61 &   -30.23 &   4961 &   4754 &   4764 &   4493 &   4449 \\
PGC 120 &   108.40 &   -37.98 &   3893 &   3771 &   3815 &   4202 &   4051 \\
PGC 129 &   108.42 &   -37.97 &   6413 &   6155 &   6105 &   4177 &   4026 \\
PGC 176 &    94.33 &   -63.84 &   7662 &   7359 &   --- &   6412 &   6117 \\
PGC 186 &   107.24 &   -42.50 &   9047 &   8557 &   9158 &   8034 &   7541 \\
PGC 195 &   355.68 &   -77.39 &   6281 &   5862 &   6319 &   6254 &   6538 \\
\hline
\end{tabular}\\
\label{TabDis}
\end{table*}

\begin{figure}
\includegraphics[width=8.5cm]{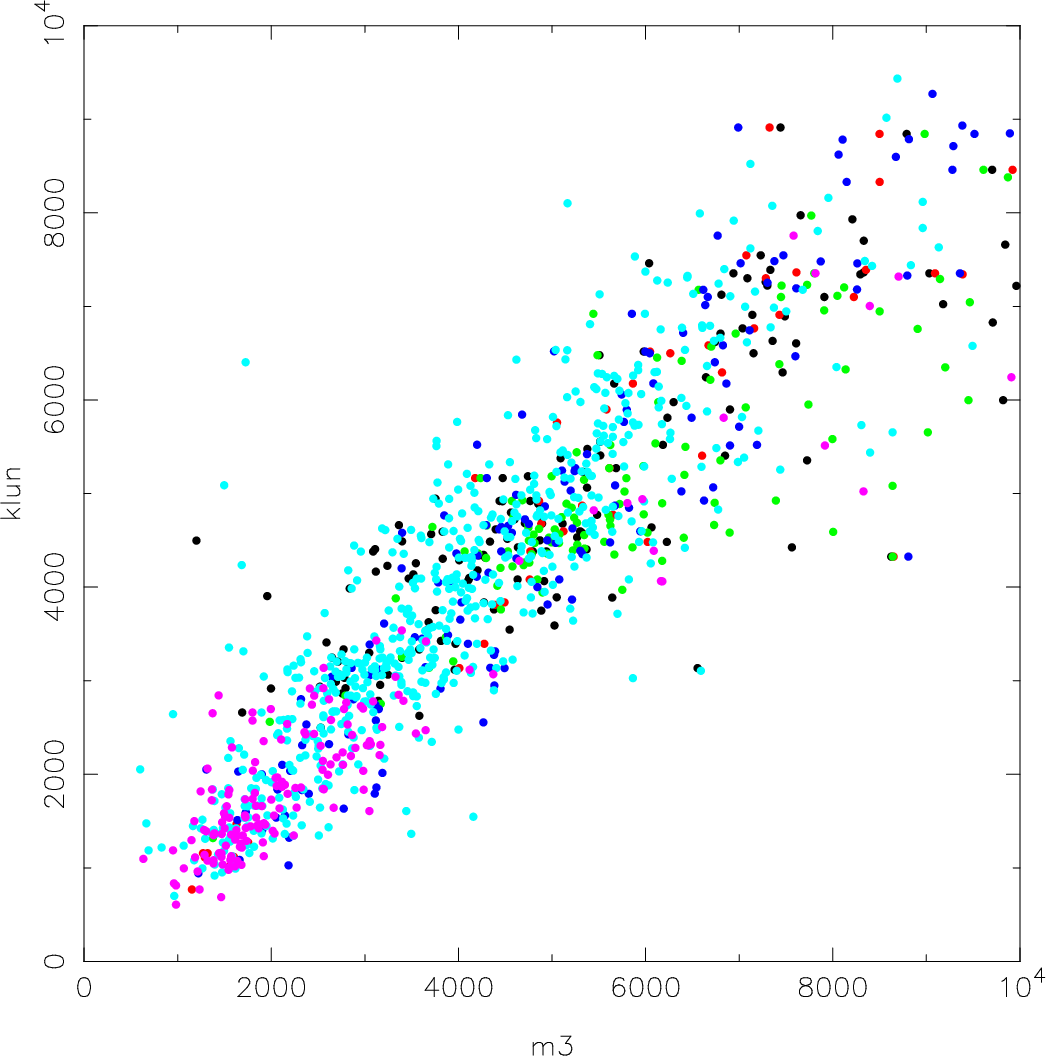}
\caption{Comparison of MarkIII and KLUN distances expressed in km.s$^{-1}$.
The colors represent the different Mark III samples (Willick et al. \cite{wil97}): 
HMCL is black, W91CL red, W91PP green, CF blue, MAT turquoise, and A82 purple.}
\label{FigMark}
\end{figure}

Our peculiar velocities were obtained for all points in space
having large enough galaxy density. We required
that there should be more than 15 galaxies with peculiar velocity measurements
within the smoothing radius around the point. Then we fit the
9-parameter tensor field to the peculiar velocities of these galaxies
and set the value obtained at the center of the smoothing window
(see Sect. 4 for more details).

Since we use a distance dependent smoothing radius, the points close the
Local Group must have a higher density of KLUN galaxies around them than
the more distant points, for a succesful velocity field determination.
This explains why some of the more distant grid points have a set value
while there are apparently no galaxies around them.

We compared the data to the MarkIII distances  (Willick et al. \cite{wil97}).
Mark III catalog was compiled from six samples of TF and one of elliptical
galaxies. It was converted to a common system by adjusting
the zero points of the distance indicators. For the Malmquist bias correction
the authors reconstructed the galaxy density field from the IRAS 1.2 Jy
survey, and used it for the inhomogeneous correction formula.
The corrected distances for 2898 spirals and 544 ellipticals are
publically available and make a good comparison point for other peculiar
velocity studies.

Figure \ref{FigMark} show the Malmquist corrected distances of
individual galaxies,
as measured in Mark III, versus the corresponding value derived by us.
The relative scatter $\sigma_d / d$ is $\sim$ 0.2, correponding to an absolute
uncertainty 0.43 in magnitude scale, is fully compatible with the measured TF scatter
(see Table \ref{TabTF}).                                                                                              
A few points in Figure \ref{FigMark} show a clear mismatch. We found that these
large discrepancies are due to errors in the input data in Mark III.
These errors are listed and discussed in  Appendix \ref{AppM3E}.

Figure \ref{FigVpK} shows the radial peculiar velocity field in the supergalactic plane,
averaged over a disc having a thickness that increases towards the
edge. The thickness of the disc in the center is zero, and its opening angle is $15\degr$,
so that at the edge (at 80 h$^{-1}$ Mpc) the disk width is about 20 h$^{-1}$ Mpc.
The blue colors refer to regions where the radial peculiar velocity
is towards us, the red regions are outfalling. The shade of the color
corresponds to the amplitude of the motion, and is saturated at
1\,000~km~s$^{-1}$. The regions where the threshold requirement of at least
two velocity measurements is not satisfied are set white.
The black dots are galaxies in HYPERLEDA with measured
redshifts (not just the KLUN galaxies). Green circles mark some well
known clusters. The maps are in real space coordinates, i.e.\ the
redshift distances corrected by the peculiar velocity field.
                                                                                                                      
It is worth noting that in our $v_{\rm pec}$ maps one observes both the front 
and backside infall patterns around the main
superclusters and structures. It is particularly obvious on Fig.\ \ref{FigVpK} for the regions of Virgo,
Perseus-Pisces, N533, Norma, or even Coma, though it is located close to the limit of the sample.
It seems that we even detect an outflow in the front side of the Great Wall. Similar features
are seen on Fig. \ref{FigSlc} that shows other slices with different orientations in space.

A wide region however, roughly centered on Centaurus cluster, seems to move away from us at a
coherent speed of $\sim$400 km\,s$^{-1}$ on a scale greater than $20h^{-1}$~Mpc. The direction
and amplitude of this bulk motion are close to the one of the putative Great Attractor 
(Lynden-Bell et al \cite{lyn88}, Hudson et al \cite{hud04}, Radburn-Smith \cite{rad06})
and cannot be associated to any structure in particular. Anyway,
it seems that this flow vanishes beyond a distance of $50h^{-1}$~Mpc.
                                  
\begin{figure*}
\includegraphics[width=16cm]{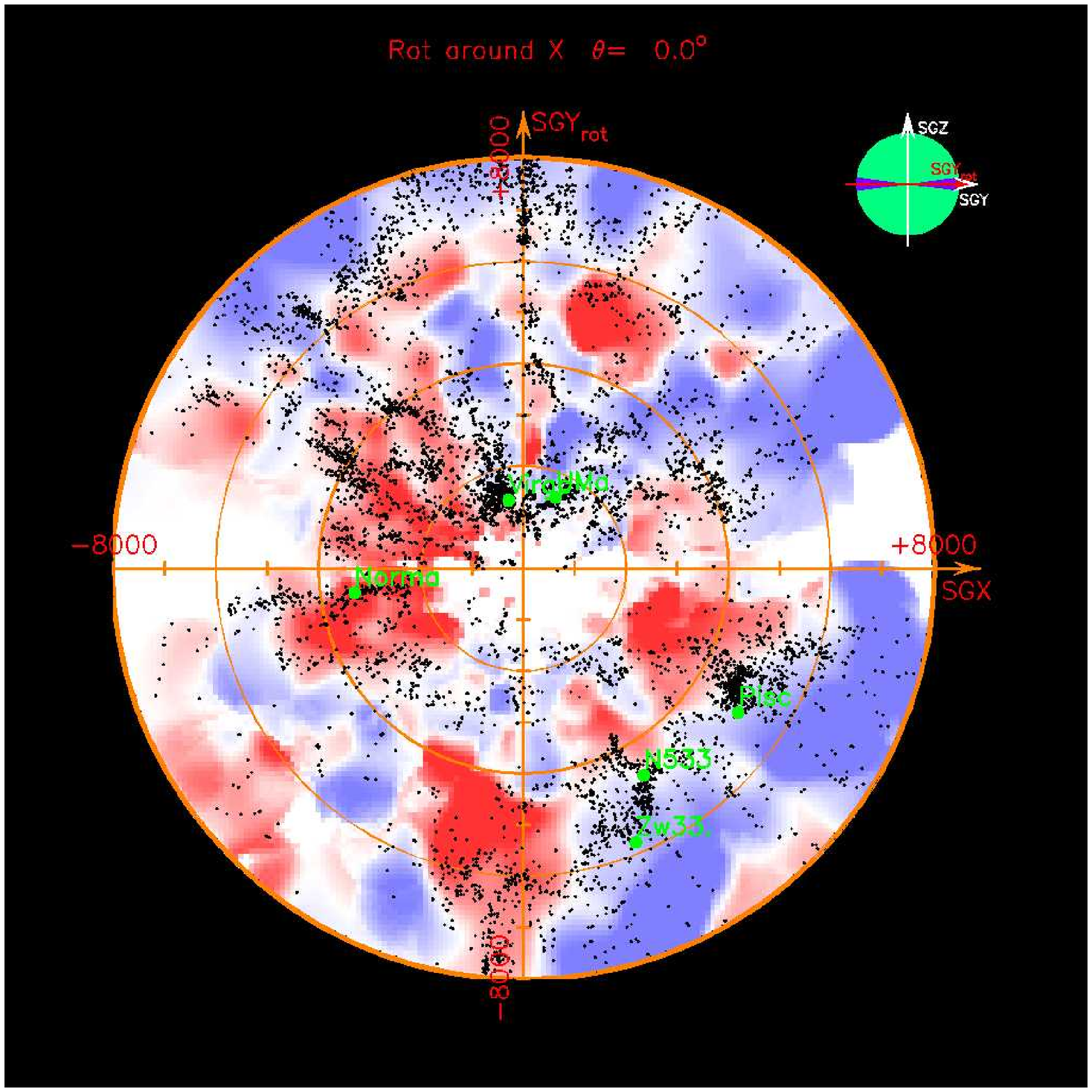}
\caption{Radial peculiar velocity field in the supergalactic plane.
Blue colors are regions with negative peculiar velocities,
red colors refer to positive ones. Black dots are all galaxies with a known
redshift. Green marks show positions of some well known structures --
here Virgo, Ursa Major, Norma, Pisces, N533, and Zwicky33. The coordinates are
in ``real space'', i.e.\ redshift distances corrected for the smoothed peculiar
velocity field, in units of km~s$^{-1}$.
}
\label{FigVpK}
\end{figure*}
                                                                                    
\begin{figure*}
\includegraphics[width=16cm]{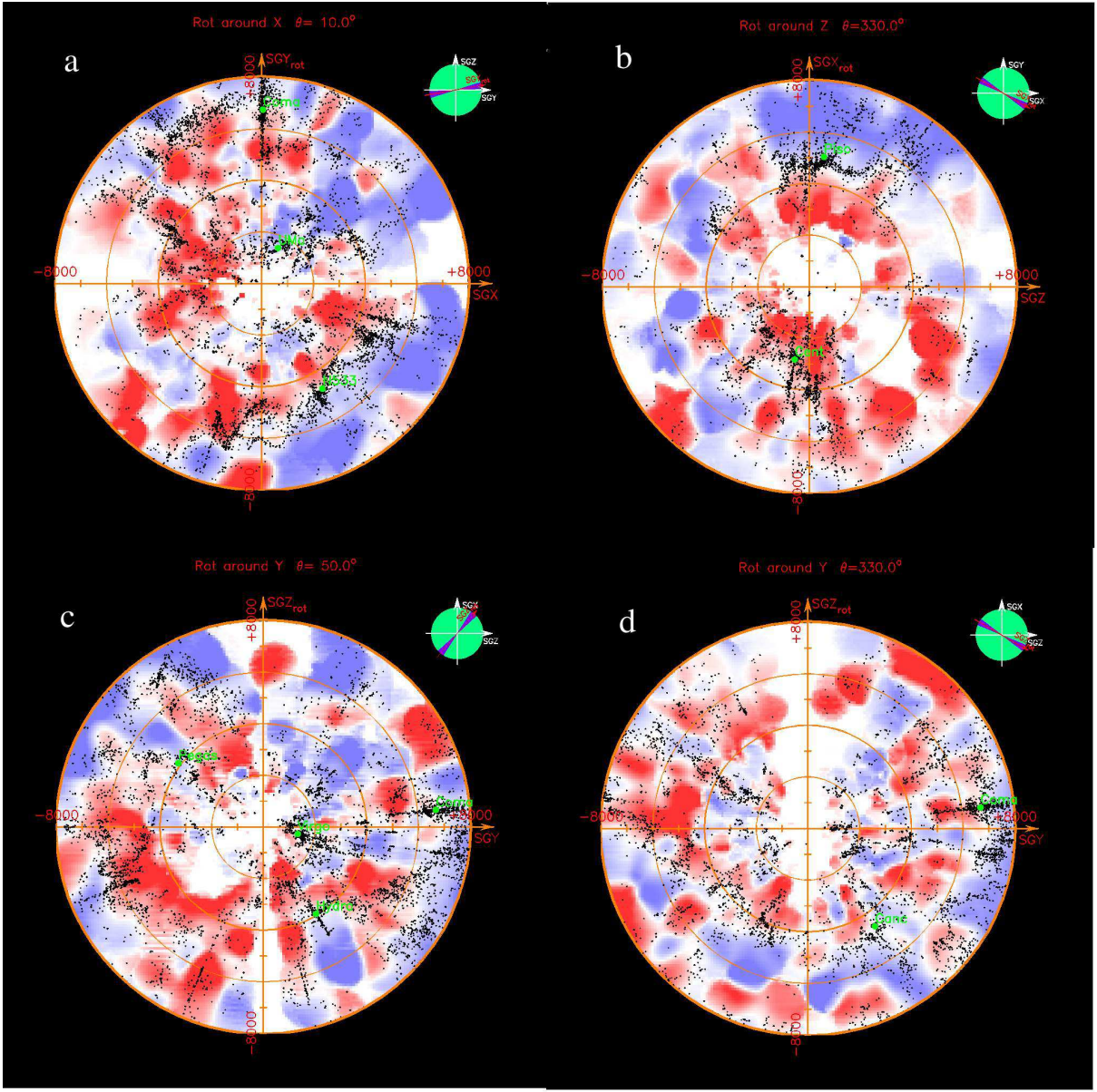}
\caption{The peculiar velocity maps, as in Fig. \ref{FigVpK},
projected on four discs of different
orientations with respect to the supergalactic plane (SGZ=0).
{\bf Figure a} is rotated by 10\degr around the SGX axis. It illustrates
well the features around two large rosary structures: the Great Wall
starting from Coma and extending counterclockwise up to the Centaurus region,
and another starting from Perseus-Pisces complex, going down
through N533 and beyond. Both front and backside infall are visible
all along the structures.
{\bf Figure b} is rotated by -30\degr around SGZ from the SGY=0 plane,
and shows the large infall
motions towards the two opposite regions, one in Perseus-Pisces and
the other in the Great Attractor or Centaurus area.
{\bf Figure c}, rotated
50\degr around SGY from the SGX=0 plane,
shows the infall patterns towards Pegasus, Hydra,
and Coma.
{\bf Figure d},
that is almost perpendicular to the supergalactic plane, rotated
-30\degr around SGY from the SGX=0 plane,
shows an example of some very detailed structure
of the velocity field.  }
\label{FigSlc}
\end{figure*}

\begin{figure}
\includegraphics[width=8.5cm,angle=270]{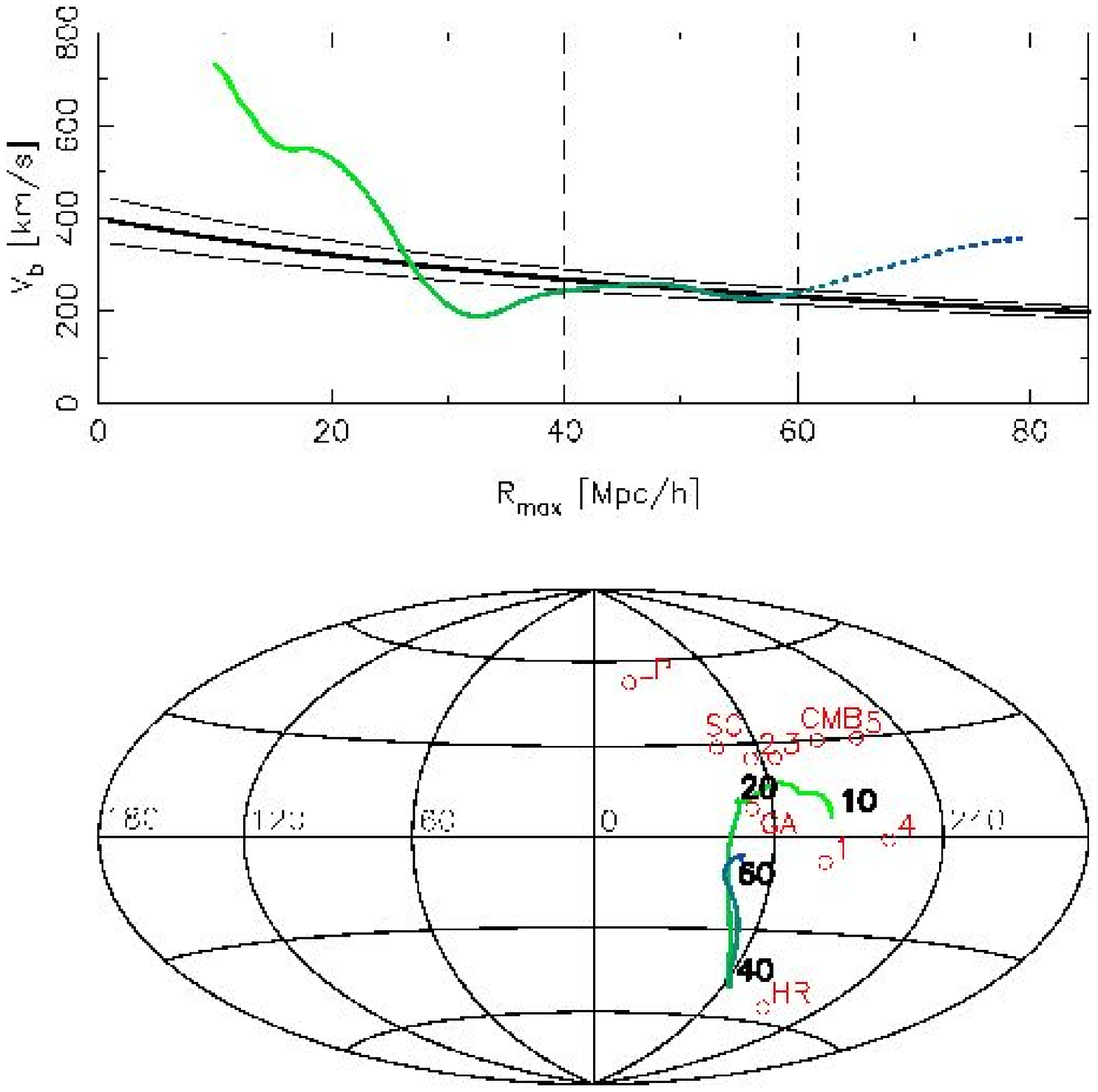}
\caption{Upper panel shows the amplitude of the observed bulk motion within
a growing sphere of radius $R_{\rm max}$, and the corresponding $\Lambda$CDM expected
curve for a flat cosmological model using : 
$n=1$, $\Omega_b = 0.04$, $h = 0.57$, and $\Omega_m$ = 0.22 $\pm$ 0.02 .
The bottom panel show the direction of the bulk flow, in galactic
coordinates, and its evolution with $R_{\rm max}=10$--$60h^{-1}$~Mpc. The main published
results as listed in Table \ref{TabBlk} are also shown.}
\label{FigBlk}
\end{figure}

As an example of quantitative result, we checked the amplitude and direction of a bulk flow within a
growing sphere centered on the Local Group. The result is shown on Fig. \ref{FigBlk}.

At short scales the direction of the flow is compatible with most previous
studies (Table \ref{TabBlk}, Fig. \ref{FigBlk}, bottom panel). In particular it coincides with the
Great Attractor for $R\sim 20h^{-1}$~Mpc. At larger scales it first drifts towards the direction of
the rich cluster region of Horologium--Reticulum, and after $R\sim 40h^{-1}$~Mpc back to
$(l,b)\sim (310\degr,-8\degr)$.

In the upper panel, we show the bulk flow amplitude within a sphere of growing radius (colored line).
It oscillates strongly at short scales, as a consequence of the density heterogeneity, and decreases 
to $\sim 250$~km~s$^{-1}$ at 40~h$^{-1}$~Mpc. Beyond this point it behaves more smoothly, as an 
indication that we reached the scale of the largest structures within our sample. It starts
to rise (or oscillate) again beyond 60~h$^{-1}$~Mpc, probably due to the sparser space sampling. 

The black solid curve shows the rms expected bulk velocity infered from the standard $\Lambda$CDM model:
\[ V_b^{rms} = \langle v^2(R) \rangle^{1/2} =
\left( \frac{\Omega_m^{1.2}}{2\pi^2} \int_0^{\infty} P(k)\tilde{W}^2(kR) dk \right)^{1/2} , \]
where $P(k)$ is the mass fluctuation power spectrum and $\tilde{W}^2(kR)$ is the Fourier transform
of a top hat window of radius $R$.
For the parametric form of $P(k)$ in the linear regime, we use the general CDM model (see e.g.
Silberman et al., \cite{sil01}) :
\[ P(k) = A_c(\Omega_m,\Omega_{\Lambda},n) T^2(\Omega_m,\Omega_b,h;k) k^n \]
$A_c$ is the normalization factor and the transfer function $T(k)$ is the one proposed by Sugiyama
(\cite{sug95}).  We restricted the analysis to a flat cosmological model with
$\Omega_m + \Omega_{\Lambda} = 1$, a scale-invariant power spectrum ($n=1$), a baryonic density
$\Omega_b = 0.04$ (WMAP result, e.g. Spergel \cite{spe06}), a Hubble constant fixed 
at $h = 0.57$ (which is our own preferred value, inferred directly from primary calibration 
by Theureau et al \cite{the97})\footnote{Note that by construction the peculiar velocities 
themselves are independent of the choice of $H_0$}, adjusting only the value of $\Omega_m$. 

The best fit has been obtained for $\Omega_m = 0.22$ in the distance range 40-60 Mpc, 
where the value of $\langle V_{rms}(R) \rangle$ appears very smoothed. We also ploted 
the $\pm$ 0.02 curves around this best value.  What we observe confirms WMAP results
on $\Omega_m$ (e.g. Spergel \cite{spe06}) and seems coherent with 
the expected rms bulk velocity within a sphere for standard $\Lambda$CDM model 
(see e.g. Willick \cite{wil00} or Zaroubi \cite{zar02}), thus with no bulk motion. 

One should be prudent anyway in such kind of conclusion : the theoretical prediction
is here the rms value of a quantity that exhibits a Maxwell distribution (see e.g. Strauss \cite{str97});
a single measurement of the flow field is only one realization out of this distribution and gives only
very weak constraints on the cosmological model.

\begin{center}
\begin{table*}
\caption{Directions, in Galactic coordinates, of some of the main bulk flow measurements
or large galaxy concentrations. The number corresponding to Fig. \ref{FigBlk}
is given in parentheses.}
\begin{tabular}{lrrl}
\hline
&$l$&$b$&ref.\\
\hline
SNIa (1)          &  282&     -8&    Riess et al. \cite{rie97}\\
ENEAR (2)         &  304&     25&    da Costa et al. \cite{dac00a}\\
SFI+SCI+II (3)    &  295&     25&    Dale \& Giovanelli \cite{dal00}\\
SMAC (4)         &  260&     -1&    Hudson et al. \cite{hud99}\\
PSCz (5)         &  260&     30&    Rowan-Robinson et al. \cite{row00}\\
LP            &  343&     52&    Lauer \& Postman \cite{lau94}\\
CMB           &  276&     30&    Local Group motion, Kogut et al. \cite{kog93}\\
GA            &  307&      9&    Great Attractor, Lynden-Bell et al. \cite{lyn88} \\
HR            &  270&    -55&    Horologium--Reticulum,  Lucey et al. \cite{luc83}\\
SC            &  315&     29&    Shapley Concentration, Scaramella et al. \cite{sca89}\\
\hline
\end{tabular}
\label{TabBlk}
\end{table*}
\end{center}

\begin{acknowledgements}
We have made use of data from the Lyon-Meudon Extragalactic
Database (HYPERLEDA). We warmly thank the scientific 
and technical staff of the Nan\c cay radiotelescope.
\end{acknowledgements}

\appendix
                                                                                                                                      
\section{Mark III errata \& rejections \label{AppM3E}}
                                                                                                                                      
When closely inspecting the Mark III data we found a few inaccuracies.
In comparison we used
the Mark III catalog provided by the CDS archives,
http://cdsweb.u-strasbg.fr, cat.\ VII/198, and the data
given by the HYPERLEDA database, http://leda.univ-lyon1.fr, as they were presented
in May 2003. A few values in Mark III were replaced by those listed
in HYPERLEDA. Some of the Mark III galaxies were rejected, having large
differences to the HYPERLEDA values.
                                                                                                                                      
Table \ref{TabApgc} lists galaxies having their PGC numbers incorrectly identified
in Mark III. We list first the number given in Mark III, followed by the
correct number, alternative name, and the Mark III data set including the
galaxy.
\begin{table*}
\caption{Mark III errata: PGC numbers.}
\begin{tabular}{llll}
\hline
Mark III & correct & alt.\ name & Mark III data sets  \\
\hline
PGC 10631$^1$ & PGC 95735 & --- & HMCL, W91CL, CF \\
PGC 64575 & PGC 64632 & NGC 6902 &  HMCL \\
PGC 57053 & PGC 57058 & UGC 10186 & W91CL \\
PGC 71291 & PGC 71292 & UGC 12572 & W91CL \\
\hline
\end{tabular}\\[1mm]
1) PGC 10631 = UGC 2285 is projected on PGC 95735, but at lower redshift.
Mark III clearly refers to the latter, but lists it as UGC 2285.
\label{TabApgc}
\end{table*}

Table \ref{TabAv} lists galaxies that were rejected due to their suspicious values
for redshift velocities (values given in the CMB rest frame). All the
galaxies with $|cz_{\mathrm M3}-cz_{\mathrm HL}|>200$~km~s$^{-1}$ were studied,
for most we maintained the Mark III values.
\begin{table*}
\caption{Mark III errata: velocities.}
\begin{tabular}{lrrl}
\hline
\hspace*{\stretch{1}} & Mark III & HYPERLEDA & Mark III data sets \\
\hline
PGC 26680 & 7266 & 12736 & HMCL \\
PGC 72301 & 11747 & 7348 & W91PP \\
PGC 17136$^1$ & 4417 & 8870 & CF \\
PGC 26561$^2$ & 2136 & 1664 & MAT \\
PGC 67258 & 2278 & 2684 & MAT \\
PGC 62889 & 2513 & 5745 & MAT \\
PGC 30753$^3$ & 2946 & 3813 & MAT \\
PGC 9551  & 4057 & 4633 & MAT \\
PGC 20324$^4$ & 4114 & 5681 & MAT \\
PGC 15790 & 4202 & 6199 & MAT \\
PGC 47832 & 4477 & 4827 & MAT \\
PGC 31723$^5$ & 4866 & 4130 & MAT \\
PGC 62411$^6$ & 5075 & 5996 & MAT \\
PGC 5964  & 5859 & 5406 & MAT \\
PGC 3144  & 5974 & 5053 & MAT \\
PGC 8888  & 6070 & 5049 & MAT \\
PGC 64523 & 6384 & 5271 & MAT \\
PGC 19363 & 7092 & 6666 & MAT \\
PGC 2001  & 7112 & 6180 & MAT \\
PGC 44349 & 9696 & 9916 & MAT \\
\hline
\end{tabular}\\[1mm]
1) Value incorrectly copied from the data source\\
2) Lower value suggested by three independent sources\\
3) See note at Giovanelli et al. \cite{gio97} (Antlia 27146)\\
4) Possible confusion with galaxy 1' SE\\
5) A typographical error in MAT file? See Mathewson et al. \cite{mat92}, Fig.\ 3\\
6) Unclear H$\alpha$ observation\\
\label{TabAv}
\end{table*}

Table \ref{TabAlogvm} has the galaxies rejected due to the $\log V_{\rm m}$ uncertainties.
Here we considered galaxies with $|\log V_{\rm m,M3}-
\log V_{\rm m,HL}|>0.15$. Notice that Mark III $\log W$ values are here
converted to the inclination corrected $\log V_{\rm m}$ of HYPERLEDA.
\begin{table*}
\caption{Mark III errata: $\log V_{\rm m}$.}
\begin{tabular}{llll}
\hline
\hspace*{\stretch{1}} & Mark III & HYPERLEDA & Mark III data sets \\
\hline
PGC 72024 & 2.321 & 2.131 & HMCL,W91CL,W91PP,CF \\
PGC 47707$^{1,2}$ & 2.139 & 2.391 & HMCL \\
PGC 65338$^1$ & 2.079 & 1.844 & HMCL \\
PGC 7111$^1$ & 1.978 & 2.251 & HMCL \\
PGC 36875$^1$ & 2.253 & 2.453 & W91CL \\
PGC 9841$^3$ & 2.049 & 1.596 & W91PP \\
PGC 51784 & 2.209 & 1.889 & CF \\
PGC 5453 & 2.429 & 1.980 & CF \\
PGC 17113$^1$ & 1.994 & 1.830 & MAT \\
PGC 16201 & 1.807 & 2.092 & MAT \\
PGC 32821$^1$ & 2.023 & 1.861 & MAT \\
PGC 67818$^1$ & 2.052 & 1.843 & MAT \\
PGC 39139 & 1.760 & 2.018 & MAT \\
PGC 51982$^1$ & 2.284 & 2.057 & MAT \\
PGC 24328 & 2.232 & 2.013 & MAT \\
PGC 64042$^1$ & 2.442 & 2.200 & MAT \\
PGC 13778$^1$ & 2.295 & 2.123 & MAT \\
PGC 2383 & 1.978 & 2.150 & MAT \\
PGC 36875$^1$ & 2.241 & 2.453 & A82 \\
\hline
\end{tabular}\\[1mm]
1) These galaxies have large differences in inclination stated in Mark III
and in HYPERLEDA.\\
2) Mark III states the inclination as $90\degr$, while it is $34\degr$
in HYPERLEDA. The latter is probably correct.\\
3) Several sources favour the HYPERLEDA value.\\
\label{TabAlogvm}
\end{table*}


\end{document}